\documentclass[11pt, a4paper]{article}

\usepackage[utf8]{inputenc}
\usepackage[T1]{fontenc}
\usepackage{lmodern}

\usepackage[margin=1in]{geometry}
\usepackage{setspace}
\usepackage{titlesec}
\titleformat*{\section}{\large\bfseries}
\titleformat*{\subsection}{\normalsize\bfseries}
\usepackage{amsmath, amssymb, amsthm, amsfonts, mathtools, bm}
\allowdisplaybreaks

\usepackage[colorinlistoftodos, textsize=footnotesize]{todonotes}
\newcounter{DannyComment}

\usepackage[authoryear,round]{natbib}

\definecolor{darkblue}{rgb}{0.0, 0.0, 0.55}
\usepackage[pagebackref=true, colorlinks=true, allcolors=darkblue]{hyperref}
\renewcommand*{\backref}[1]{}
\renewcommand*{\backrefalt}[4]{%
    \ifcase #1%
    \or #2%
    \else #2%
    \fi%
}

\makeatletter
\newtheoremstyle{smallcaps}
  {}
  {}
  {\itshape}
  {}
  {\scshape}
  {.}
  { }
  {}

\renewenvironment{proof}[1][\proofname]{\par
  \pushQED{\qed}%
  \normalfont \topsep6\p@\@plus6\p@\relax
  \trivlist
  \item[\hskip\labelsep
        \scshape
    #1\@addpunct{.}]\ignorespaces
}{%
  \popQED\endtrivlist\@endpefalse
}
\makeatother
\theoremstyle{smallcaps}
\newtheorem{theorem}{Theorem}[section]
\newtheorem{lemma}[theorem]{Lemma}

\newtheorem{claim}[theorem]{Claim}

\newcommand{\MyAbove}[2]{\genfrac{}{}{0pt}{}{#1}{#2}}
\newcommand{\eps}{\epsilon}
\newcommand{\bbR}{\mathbb{R}}
\newcommand{\bbN}{\mathbb{N}}
\newcommand{\myeoq}{\mathrm{EOQ}}
\newcommand{\opt}{\mathrm{OPT}}
\newcommand{\mycyc}{\mathrm{cycle}}
\newcommand{\mydp}{\mathrm{DP}}
\newcommand{\mylb}{\mathrm{L}}
\newcommand{\myub}{\mathrm{U}}
\newcommand{\mynext}{\mathrm{next}}

\title{\textbf{Economic Warehouse Lot Scheduling: \\
Approximation Schemes via \\
Efficiently-Representable DP-Encoded Policies}}
\author{
    Danny Segev\thanks{School of Mathematical Sciences and Coller School of Management, Tel Aviv University, Tel Aviv 69978, Israel. Email: \texttt{segevdanny@tauex.tau.ac.il}. Supported by Israel Science Foundation grant 1407/20.}
}
\date{}

\begin{document}

\maketitle

\begin{abstract}
\noindent In this focused technical paper, we present long-awaited algorithmic advances toward the efficient construction of near-optimal replenishment policies for a true inventory management classic, the economic warehouse lot scheduling problem. While this paradigm has accumulated a massive body of surrounding literature since its inception in the late '50s, we are still very much in the dark as far as basic computational questions are concerned, perhaps due to the intrinsic complexity of dynamic policies in this context. The latter feature forced earlier attempts to either study highly-structured classes of policies or to forgo provably-good performance guarantees altogether; to this day, rigorously analyzable results have been few and far between.

\medskip \noindent The current paper develops novel analytical foundations for directly competing against dynamic policies. Combined with further algorithmic progress and newly-gained insights, these ideas culminate in a polynomial-time approximation scheme for constantly-many commodities. In this regard, the efficient design of $\eps$-optimal dynamic policies appeared to have been out of reach, since beyond their inherent algorithmic challenges, even the polynomial-space representation of such policies has been a fundamental open question.
\end{abstract}

\vspace{1em}

\thispagestyle{empty}

\newpage
\thispagestyle{empty}
\tableofcontents

\newpage
\setcounter{page}{1}

\section{Introduction}

The primary intent of this focused technical paper is to present long-awaited algorithmic advances toward the efficient construction of near-optimal replenishment policies for a true inventory management classic, the economic warehouse lot scheduling problem. Appearing since the late '50s under various names, decades-long investigations of this fundamental paradigm have led to a massive body of surrounding literature. However, we are still very much in the dark as far as basic computational questions are concerned, perhaps due to the inherent complexity of dynamic replenishment policies in this context, forcing earlier attempts to either study highly-structured classes of policies or to forgo provably-good performance guarantees altogether. To this day, rigorously analyzable results have been few and far between. The current paper develops novel analytical foundations for directly competing against dynamic policies; combined with further algorithmic progress and newly-gained insights, these ideas culminate in a polynomial-time approximation scheme for constantly-many commodities. As explained later on, the efficient design of $\eps$-optimal dynamic policies appeared to have been out of reach, since beyond their inherent algorithmic challenges, even the polynomial-space representation of such policies has been a major open question. 

Given the extensive body of work that has accumulated around the economic warehouse lot scheduling problem, including heuristic approaches, experimental studies, and concrete  applications, there is no way we can do justice and present an in-depth overview of this literature. For a broader perspective on these topics, readers are referred to excellent book chapters \citep{HadleyW1963, JohnsonM74, HaxC84, Zipkin00, SimchiLeviCB14, NahmiasL15}, as well as to the references therein, noting that it may be especially interesting to revisit pioneering papers such as those of \cite{Holt58}, \cite{Homer1966}, \cite{PageP76}, and \cite{Zoller77}. From a bird's-eye view, our problem of interest and most of its nearby relatives are all intrinsically concerned with  lot-sizing and scheduling decisions for multiple commodities over a given planning horizon, where we wish to minimize long-run average operating costs. However, on top of marginal ordering and inventory holding costs, a particularly problematic feature of this model is the interaction between different commodities, incurred due to sharing a common resource; the latter will be referred to as ``warehouse space'', just to align ourselves with classic terminology. In this regard, capacity-feasible replenishment policies should ensure that, at any point in time, the momentary inventory levels of all commodities can be jointly packed within our warehouse's capacity, with each commodity contributing toward this space requirement at an individual rate. Unfortunately, this seemingly-innocent constraint generally forces optimal policies to violate many fundamental properties that have been explored and exploited for single-commodity models. Consequently, we have very sparse theoretical grounds to rely on, leading to long-standing algorithmic questions about how provably-good capacity-feasible policies can be efficiently identified, as well as to a deep analytical void regarding their structural characterization.

Moving forward, to delve into the finer details of these questions, Section~\ref{subsec:model_definition} provides a complete mathematical description of the economic warehouse lot scheduling problem in its broadest  form. Subsequently, Section~\ref{subsec:related_work} sheds light on cornerstone theoretical advances, including the well-known SOSI-driven approach, whose approximation guarantees have been state-of-the-art to this day. Concurrently, we will elaborate on the primary open questions that motivated our work. Finally, Section~\ref{subsec:contributions} discusses the main contributions of this paper, leaving their structural characterization, algorithmic ideas, and analytical arguments
to be presented in subsequent sections.

\subsection{Model formulation} \label{subsec:model_definition}

\paragraph{The economic order quantity model.} To describe the inner workings of our model in an accessible way, we begin by introducing its basic building block, the economic order quantity (EOQ) model. At a high level, our objective is to identify an optimal dynamic replenishment policy for a single commodity, aiming to minimize its long-run average cost over the continuous planning horizon $[0,\infty)$. Without loss of generality, this commodity is assumed to be characterized by a stationary demand rate of $1$, to be completely met upon occurrence, meaning that lost sales or back orders are not permitted. In this context, the notion of a ``dynamic'' policy ${\cal P}$ will be captured by two ingredients: 
\begin{itemize}
    \item A sequence of ordering points $0 = \tau_0 < \tau_1 < \tau_2 < \cdots$ covering the entire planning horizon $[0,\infty)$, in the sense that $\lim_{k \to \infty} \tau_k = \infty$.

    \item The respective real-valued ordering quantities $q_0, q_1, q_2, \ldots$ at these points.
\end{itemize}
For ease of notation, $I( {\cal P}, t )$ will stand for the inventory level we end up with at time $t$, when operating under the replenishment policy ${\cal P}$. One straightforward way to express this function is via the relation $I( {\cal P}, t ) = \sum_{k \geq 0: \tau_k \leq t} q_k - t$. To parse this expression, note that $\sum_{k \geq 0: \tau_k \leq t} q_k$ represents the total ordering quantity up to and including time $t$; this term consists of finitely-many summands, since $\lim_{k \to \infty} \tau_k = \infty$. In addition, the second term, $t$, is precisely the cumulative consumption up until this time, due to having a unit demand rate. As such, we are indeed meeting our demand at any point in time via previously-placed orders if and only if $I( {\cal P}, t ) \geq 0$ for all $t \in [0,\infty)$. Any policy satisfying this condition is said to be feasible, and we designate the collection of such policies by ${\cal F}$.

We proceed by explaining how the long-run average cost of any policy ${\cal P} \in {\cal F}$ is structured. For this purpose, each of the ordering points $\tau_0, \tau_1 ,\tau_2, \ldots$ incurs a fixed set-up cost of $K$, regardless of its quantity. In other words, letting $N( {\cal P}, [0,t)) = | \{ k \in \bbN_0 : \tau_k < t \} |$ be the number of orders placed along $[0,t)$, this interval is associated with a total ordering cost of ${\cal K}( {\cal P}, [0,t)) = K \cdot N( {\cal P}, [0,t))$. Intuitively, these costs by themselves motivate us to place infrequent orders. On the other hand, we are concurrently facing a linear holding cost of $2H$, incurred per time unit for each inventory unit in stock. Technically speaking, ${\cal H}( {\cal P}, [0,t))$ will denote our total holding cost across the interval $[0,t)$, given by ${\cal H}( {\cal P}, [0,t)) = 2H \cdot \int_{[0,t)} I( {\cal P}, \tau ) \mathrm{d} \tau$. These costs by themselves are pulling us in the opposite direction, as we generally wish to avoid high inventory levels via frequent orders. Putting both ingredients into a single objective, the combined cost of a feasible policy ${\cal P}$ along $[0,t)$ will be designated by $C( {\cal P}, [0,t)) = {\cal K}( {\cal P}, [0,t)) + {\cal H}( {\cal P}, [0,t))$. In turn, the long-run average cost of this policy can be specified by
\begin{equation} \label{eqn:long_run_cost_single}
C( {\cal P} ) ~~=~~ \limsup_{t \to \infty} \frac{ C( {\cal P}, [0,t)) }{ t } \ .    
\end{equation}
As a side note, one can easily come up with feasible policies ${\cal P}$ for which $\lim_{t \to \infty} \frac{ C( {\cal P}, [0,t)) }{ t }$ does not exist, meaning that the $\limsup$ above is indeed required.

Based on the preceding description, in the economic order quantity problem, our goal is to identify a feasible replenishment policy ${\cal P}$ whose long-run average cost is minimized, in the sense that $C( {\cal P} ) = \min_{ \hat{\cal P} \in {\cal F}} C( \hat{\cal P} )$. By consulting  relevant textbooks, such as those of \citet[Sec.~3]{Zipkin00}, \citet[Sec.~2]{MuckstadtS10}, or \citet[Sec.~7.1]{SimchiLeviCB14}, readers will note that the latter minimum is indeed attained, via the  extremely simple class of ``stationary order sizes and stationary intervals'' (SOSI) policies. These policies are characterized by a single parameter, $T$, standing for the uniform time interval between successive orders. Having fixed this parameter, orders will be placed at the time points $0, T, 2T, 3T, \ldots$, each consisting of exactly $T$ units, meaning that zero-inventory levels are reached at each of these points. As such, since the long-run average cost of this policy can be written as $C_{\myeoq}(T) = \frac{ K }{ T } + HT$, the economic order quantity problem admits an elegant closed-form solution. The next claim summarizes several well-known properties exhibited by this function, all following from elementary
calculus arguments.

\begin{claim} \label{clm:EOQ_properties}
The cost function $C_{\myeoq}(T) = \frac{ K }{ T } + HT$ satisfies the next few properties:
\begin{enumerate}
    \item $C_{\myeoq}$ is strictly convex. 

    \item The unique minimizer of $C_{\myeoq}$ is $T^* = \sqrt{ K / H }$. 

    \item $C_{\myeoq}( \alpha T ) \leq \max \{ \alpha, \frac{ 1 }{ \alpha } \} \cdot C_{\myeoq}(T)$, for every $\alpha > 0$ and $T > 0$.  
    \end{enumerate}
\end{claim}

\paragraph{The economic warehouse lot scheduling problem.} Given  the optimality of SOSI policies for the economic order quantity model, one can retrospectively wonder whether placing dynamic policies at the heart of our exposition makes sense. Returning to this issue in the sequel, we proceed by explaining how these foundations allow for a smooth transition to the economic warehouse lot scheduling problem, whose essence can be encapsulated into the next high-level question:
\begin{quote}
{\em How should we coordinate multiple economic order quantity models, when different commodities are tied together by sharing a common resource?}
\end{quote}
Specifically, we wish to synchronize the lot sizing of $n$ distinct commodities, where each commodity $i \in [n]$ is coupled with its own EOQ model, parameterized by ordering
and holding costs $K_i$ and $2H_i$, respectively. As explained above, deciding to replenish this commodity via the dynamic policy ${\cal P}_i \in {\cal F}$ would lead to a marginal long-run average cost of $C_i( {\cal P}_i )$, prescribed by equation~\eqref{eqn:long_run_cost_single}. However, the complicating caveat is that we are concurrently facing a ``warehouse space'' constraint, stating that the momentary inventory levels of all commodities can be jointly packed within our warehouse's capacity, with each commodity contributing toward this space requirement at an individual rate. 

To formalize this notion, for every commodity $i \in [n]$, each inventory unit requires $\gamma_i$ amount of space to be stocked in a common warehouse, whose overall capacity will be denoted by ${\cal V}$. Consequently, for any dynamic replenishment policy ${\cal P} = ( {\cal P}_1, \ldots, {\cal P}_n ) \in {\cal F}^n$, by recalling that $\{ I( {\cal P}_i, t) \}_{i \in [n]}$ designate its underlying inventory levels at time $t \in [0, \infty)$, it follows that the  warehouse space occupied at that point in time is given by $V( {\cal P}, t) = \sum_{i \in [n]} \gamma_i \cdot I( {\cal P}_i, t )$. In turn, the peak space ever occupied by this policy can be written as $V_{\max}( {\cal P} ) = \sup_{t \in [0, \infty)}  V( {\cal P}, t)$, and we say that ${\cal P}$ is capacity-feasible when $V_{\max}( {\cal P} ) \leq {\cal V}$. Clearly, this condition is much stronger than merely asking each of the marginal policies ${\cal P}_1, \ldots, {\cal P}_n$ to be feasible by itself.

Putting things together, in the economic warehouse lot scheduling problem, our goal is
to determine a capacity-feasible replenishment policy ${\cal P} = ( {\cal P}_1, \ldots, {\cal P}_n )$ whose long-run average
cost $C( {\cal P} ) = \sum_{i \in [n]} C_i( {\cal P}_i )$ is minimized. Moving forward, it will be convenient to succinctly formulate this model as 
\begin{align} \label{eqn:model_warehouse} \tag{$\Pi$}
\begin{array}{ll}
{\displaystyle \min_{{\cal P} \in {\cal F}^n}} & C( {\cal P} ) \\
\text{s.t.} & V_{\max}( {\cal P} ) \leq {\cal V} 
\end{array} 
\end{align}
It is worth pointing out that the above-mentioned minimum is indeed attained by some capacity-feasible policy. The proof of this claim is rather straightforward, and we leave its details to be rediscovered by avid readers.

\subsection{Known results and open questions} \label{subsec:related_work}

In what follows, we survey cornerstone results regarding the economic warehouse lot scheduling problem, with a particular emphasis on those that are directly related to our basic research questions. For this purpose, we will be going in two complementary directions, one centered around rigorous algorithmic methods for efficiently identifying provably-good replenishment policies, and the other briefly discussing known intractability results; as previously mentioned, such findings have been extremely rare. In regard to tangential research directions, given the extensive literature in this context, it is impractical to present an all-inclusive overview. Further background on historical developments, heuristic approaches, and experimental studies can be attained by consulting numerous book chapters dedicated to these topics \citep{HadleyW1963, JohnsonM74, HaxC84, Zipkin00, SimchiLeviCB14, NahmiasL15}. 

\paragraph{Hardness results.} While our work is algorithmically driven for the most part, to thoroughly understand the boundaries of provably-good performance guarantees, it is worth briefly mentioning known intractability results. Along these lines, unlike many deterministic inventory models whose computational status is very much unclear, \cite*{GallegoSS92} were  successful in rigorously examining the plausibility of efficiently computing optimal replenishment policies. Their main finding in this context consists of a polynomial-time reduction from the well-known $3$-partition problem \citep{GareyJ75, GareyJ79}, showing that economic warehouse lot scheduling is strongly NP-hard in its decision problem setup. As an immediate byproduct, we know that optimal replenishment policies cannot be computed in polynomial time unless $\mathrm{P} = \mathrm{NP}$; moreover, this setting does not admit a fully polynomial-time approximation scheme under the same complexity assumption \citep[Sec.~8.3]{Vazirani01}.

\paragraph{Representation issues.} On top of complexity-related findings in terms of computation time, a seemingly unsurpassable obstacle on the road to efficiently identifying truly near-optimal policies can be captured by the next two  questions: 
\begin{quote}
\begin{itemize}
    \item {\em Is there any way we can represent optimal or $\eps$-optimal dynamic policies in polynomially-bounded memory space?}

    \item {\em Even more basically, can we address this challenge for $O(1)$ commodities?}
\end{itemize}
\end{quote}
In other words, suppose we are not worried about running times, and instead, we are only wondering about the feasibility of efficiently describing near-optimal dynamic policies. One could attempt to tackle this fundamental issue, for example, by proving that there exists a polynomially-representable pattern behind the sequence of ordering points $\tau_0^i, \tau_1^i, \tau_2^i, \ldots$ and quantities $q_0^i, q_1^i, q_2^i, \ldots$ associated with each commodity $i \in [n]$. Alternatively, one may wish to argue about the existence of a short-duration cyclic policy, where we will be seeing only a polynomial number of ordering points. To our knowledge, these questions have been wide open for decades, even for the most basic setting of only two commodities.

\paragraph{Constant-factor approximations via SOSI policies.} Prior to diving into state-of-the-art approximation guarantees, it is worth pointing out that, by superficially browsing through the abstracts of many early papers, we could mistakenly get the impression that economic warehouse lot scheduling is well-understood, at least in stylized settings with two or three commodities. However, the intrinsic caveat is that provably-good and efficient algorithmic results along these lines are focusing on very specific classes of policies rather than considering arbitrarily-structured dynamic policies. 

The first breakthrough in this area should be attributed to the influential work of \cite{Anily91}. While  her  objective was to optimize over the class of SOSI policies, she has been able to rigorously establish a bridge between economic warehouse lot scheduling and the SOSI-restricted economic order quantity model, where the underlying commodities are interacting via a warehouse space constraint on their  peak inventory levels. As a byproduct of these ideas, Anily  devised a polynomial-time construction of a SOSI policy whose long-run average cost is within factor $2$ of the minimum-possible within this class. Subsequently, the sophisticated work of \cite*{GallegoQS96} fused Anily's approach with further insights, prescribing a lower bound on the long-run average cost of any capacity-feasible dynamic policy via an elegant convex relaxation. As a direct consequence, the authors have been successful at approximating the economic warehouse lot scheduling problem in its full generality, proposing a polynomial-time algorithm for identifying a SOSI policy whose cost is within factor $2$ of optimal; this time, ``optimal'' is referring to the best-possible dynamic policy! 

\paragraph{Main open questions.} Quite surprisingly, while we have witnessed a continuous stream of literature on this topic across the last three decades, mostly revolving around stylized settings, heuristic approaches, and experimental studies, the algorithmic ideas of \cite{Anily91} and \cite{GallegoQS96} along with their lower-bounding mechanism still represent the best-known approximation guarantees for computing dynamic replenishment policies. Moreover, at present time, there is absolutely no distinction between a constant number and an arbitrary number of commodities in terms of deriving improved guarantees. Given this state of affairs, the fundamental
questions that lie at the heart of our work, as stated in many papers, books, and course materials, can be succinctly summarized as follows: 
\begin{itemize}
    \item {\em Can we outperform these long-standing results, in any shape or form?}

    \item {\em Which algorithmic techniques and analytical ideas could be useful for this purpose?}

    \item {\em In spite of basic representation issues, can we take advantage of settings with $O(1)$ commodities for computing truly near-optimal policies?}
\end{itemize}

\subsection{Main contributions} \label{subsec:contributions}

The first and foremost contribution of this focused technical paper resides in developing long-awaited algorithmic advances, allowing us to resolve these open questions by devising a novel dynamic programming approach for computing and representing near-optimal capacity-feasible replenishment policies. Specifically, as stated in Theorem~\ref{thm:PTAS_exponential}, our $O( | {\cal I} |^{O(n)} \cdot 2^{ O( n^{6} / \eps^{5} ) } )$-time algorithm constructs an efficiently-representable cyclic policy whose long-run average cost is within factor $1 + \eps$ of optimal. While  this outcome  applies to all problem instances, it identifies with the notion of a polynomial-time approximation scheme (PTAS) for constantly-many commodities, i.e., when $n = O(1)$. 

\begin{theorem} \label{thm:PTAS_exponential}
For any $\eps > 0$, the economic warehouse lot scheduling problem can be approximated within factor $1 + \eps$ of optimal. Our algorithm admits an $O( | {\cal I} |^{O(n)} \cdot 2^{ O( n^{6} / \eps^{5} ) } )$-time implementation, where $| {\cal I} |$ stands for the input length in its binary specification.
\end{theorem}

Put differently, for $O(1)$ commodities, this result allows us to approach the best-possible long-run average cost achievable by dynamic policies within any degree of accuracy; at the same time, we are still able to  represent our near-optimal policy in polynomial space. On a different note, it is important to emphasize that we have not attempted to optimize any  running time exponents, aiming for the most accessible presentation rather than concentrating on technical minutiae. By diving into the intricacies of Sections~\ref{sec:DP-foundations} and~\ref{sec:approx_scheme_details}, expert readers will discover that such enhancements are doable, yet come at the cost of requiring tedious analysis.

Consequently, for constantly-many commodities, our work constitutes the first improvement over existing SOSI-driven approaches to economic warehouse lot scheduling, whose constant-factor guarantees have been unbeatable since the mid-90’s. Beyond this quantitative progress, we establish new-fashioned analytical foundations for directly competing against dynamic policies, which may be of broader importance and applicability. To demonstrate the viability of this claim, Section~\ref{sec:conclusions} succinctly discusses our subsequent work~\citep{Segev26EWLSPsub2}, where along with quite a few additional developments, Theorem~\ref{thm:PTAS_exponential} has been instrumental in breaking the long-standing $2$-approximation for general problem instances \citep{Anily91, GallegoQS96}, consisting of arbitrarily-many commodities.
\section{Toward a Dynamic Programming Approach} \label{sec:DP-foundations}

This section is dedicated to developing the structural foundations of our dynamic programming approach, allowing us to compute efficiently-representable near-optimal replenishment policies later on. In particular, we show that for any $\eps>0$, one can restrict attention to capacity-feasible cyclic policies whose long-run average cost is within factor $1+O(\eps)$ optimal, while admitting explicit bounds on their cycle length and on the number of orders placed for each commodity. These bounds will be central to producing a discrete space of dynamic policies that can be algorithmically handled.

\paragraph{Outline.} In Sections~\ref{subsec:structured_cyclic} and~\ref{subsec:proof_lem_good_cyclic}, we establish the existence of such policies and derive useful bounds on their cycle length and ordering structure. This result is obtained by showing how an optimal dynamic policy can be converted into being cyclic,  with controlled cost inflation and without violating capacity feasibility. Building on these insights, Sections~\ref{subsec:structure_thm} and~\ref{subsec:proof_lem_exist_aligned} introduce frequency classes and hierarchical partitions, culminating in the Alignment Theorem, which proves  the existence of efficiently-representable cyclic policies whose ordering decisions are aligned with structured breakpoint sets. Taken as a whole, these results yield  regularity properties that will be exploited in Section~\ref{sec:approx_scheme_details} to design our  approximation scheme.

\subsection{Milestone I: Existence of low-cost short-cycle policies} \label{subsec:structured_cyclic}

For the purpose of limiting our solution space to cyclic policies, we begin by investigating the following  question: How good could  cyclic policies be, when we place certain bounds on their cycle length and number of orders per commodity? As formally stated in Theorem~\ref{thm:good_cyclic} below, we prove that such policies are capable of nearly matching the long-run average cost of optimal fully-dynamic policies, with useful bounds on their cycle length and number of orders per commodity. Notably, the existence of a near-optimal cyclic policy by itself is a rather straightforward question. The crux of our approach will be to concurrently attain specific bounds on its cycle length and ordering pattern, which will be algorithmically important later on. To avoid cumbersome expressions in items~2 and~3 below, we make use of the shorthand notation ${\cal M} = \sum_{i \in [n]} ( \frac{ K_i }{ T^{\cal V}_i } + H_i T^{\cal V}_i )$, where $T^{\cal V}_i = \min \{ \sqrt{K_i/H_i}, {\cal V}/ \gamma_i \}$.

\begin{theorem} \label{thm:good_cyclic}
For any $\eps \in (0, \frac{ 1 }{ 3 } )$, there exists a capacity-feasible cyclic policy $\tilde{\cal P}$ satisfying the following properties:
\begin{enumerate}
    \item {\em Near-optimality:} $C( \tilde{\cal P} ) \leq (1 + 3\eps) \cdot \opt\eqref{eqn:model_warehouse}$.

    \item {\em Cycle length:} $\tilde{\cal P}$ has a cycle length of $\tau_{\mycyc} \in [\frac{ K_{\max} }{ 2\eps n {\cal M} }, \frac{ 2 n {\cal M} }{ \eps^2 H_{\min} }]$.

    \item {\em Orders per cycle:} $N( \tilde{\cal P}_i, [0,\tau_{\mycyc})) \in [\frac{ 1 }{ \eps }, \frac{ 4 n^2 {\cal M}^2 }{ \eps^2 K_{\min} H_{\min} }]$, for every commodity $i \in [n]$. Additionally, $N( \tilde{\cal P}_i, [0,\tau_{\mycyc})) = \frac{ 1 }{ \eps }$ for at least one of these commodities.
\end{enumerate}
\end{theorem}

We mention in passing that, while a complete proof of this result is provided in Section~\ref{subsec:proof_lem_good_cyclic}, its statement is modular, in the sense of allowing readers to proceed directly into milestone~II, whose specifics are discussed in Section~\ref{subsec:structure_thm}.

\subsection{Proof of Theorem~\ref{thm:good_cyclic}} \label{subsec:proof_lem_good_cyclic}

\paragraph{Constructing {\boldmath${\tilde{\cal P}}$}.} For the purpose of deriving this result, let ${\cal P}^*$ be an optimal policy. With respect to ${\cal P}^*$, we consider the infinite sequence $\tau_1 < \tau_2 < \cdots$ of time points, specified as follows:
\begin{itemize}
    \item First, $\tau_1$ is the minimal $t > 0$ for which the segment $[0,t]$ has: (a)~At least $\frac{ 1 }{ \eps } + 1$ orders of every commodity $i \in [n]$; and (b)~Exactly $\frac{ 1 }{ \eps } + 1$ orders of some commodity $i \in [n]$.

    \item In turn, $\tau_2$ is the minimal $t > \tau_1$ for which the segment $[\tau_1,t]$ meets conditions~(a) and~(b) above.

    \item The remaining sequence $\tau_3, \tau_4, \ldots$ is defined in a similar way.
\end{itemize}
By recalling how dynamic replenishment policies are defined (see Section~\ref{subsec:model_definition}), it is easy to verify that the sequence $\tau_1, \tau_2, \ldots$ is well-defined and that it tends to infinity.

In terms of this sequence, the optimal long-run average cost can be written as $C( {\cal P^*} ) = \lim_{\hat{\kappa} \to \infty} \frac{ C( {\cal P}^*, [0, \tau_{\hat{\kappa}} )) }{ \tau_{\hat{\kappa}} }$. Noting that
$\tau_{\hat{\kappa}} = \sum_{ \kappa \leq \hat{\kappa} } ( \tau_{\kappa} - \tau_{\kappa-1} )$ and $C( {\cal P}^*, [0, \tau_{\hat{\kappa}} )) = \sum_{ \kappa \leq \hat{\kappa} } C( {\cal P}^*, [\tau_{\kappa-1}, \tau_{\kappa} ))$, it follows that  for every $\eps >0$, there exists an index $\kappa = \kappa(\eps)$ for which
\begin{equation} \label{eqn:proof_lem_good_cyclic_eq1}
\frac{ C( {\cal P^*}, [ \tau_{\kappa}, \tau_{\kappa+1}) ) }{ \tau_{\kappa+1} - \tau_{\kappa} } ~~\leq~~ (1 + \eps) \cdot C( {\cal P^*} ) \ .
\end{equation}
As such, our cyclic policy $\tilde{\cal P}$ is created by replicating the actions of ${\cal P}^*$ over the segment $[ \tau_{\kappa}, \tau_{\kappa+1})$ across the entire planning horizon $[0,\infty)$. To align every pair of successive cycles, we decrement the last order of each commodity $i \in [n]$, thereby making sure that each  cycle ends with zero inventory. In addition, at the beginning of each cycle, we place an extra order of this commodity, consisting of $I( {\cal P}^*_i, \tau_{\kappa})$ units. From this point on, $\tau_{\mycyc} = \tau_{\kappa+1} - \tau_{\kappa}$ will denote the cycle length of this policy.

\paragraph{Feasibility and cost.} It is easy to verify that our resulting policy $\tilde{\cal P}$ is capacity-feasible, since
\[ V_{\max}( \tilde{\cal P} ) ~~\leq~~ \max_{\tau \in [ \tau_{\kappa}, \tau_{\kappa+1})}  V( {\cal P}^*, \tau)  ~~\leq~~ V_{\max}( {\cal P}^* ) ~~\leq~~ {\cal V} \ , \]
where the last inequality holds since the optimal policy ${\cal P}^*$ is in particular capacity-feasible. To upper bound the long-run average cost of $\tilde{\cal P}$, let us break this measure into its holding and ordering costs as follows:
\begin{itemize}
    \item {\em Holding}: Along a single cycle, we incur a holding cost of ${\cal H}( \tilde{\cal P}, [0,\tau_{\mycyc})) \leq {\cal H}( {\cal P}^*, [ \tau_{\kappa}, \tau_{\kappa+1}))$, due to having $I( \tilde{\cal P}_i, \tau) \leq I( {\cal P}^*_i, \tau_{\kappa} + \tau)$ for all $i \in [n]$ and $\tau \in [0, \tau_{\mycyc})$.

    \item {\em Ordering}: Along a single cycle, we have added at most one extra order for each commodity $i \in [n]$ in comparison to ${\cal P}^*$, implying that
    \[ N( \tilde{\cal P}_i, [0,\tau_{\mycyc})) ~~\leq~~ N( {\cal P}^*_i, [ \tau_{\kappa}, \tau_{\kappa+1})) + 1 ~~\leq~~ (1 + \eps) \cdot  N( {\cal P}^*_i, [ \tau_{\kappa}, \tau_{\kappa+1})) \ , \]
    where the last inequality holds since $N( {\cal P}^*_i, [ \tau_{\kappa}, \tau_{\kappa+1})) \geq \frac{ 1 }{ \eps }$, by definition of $\tau_{\kappa+1}$; see condition~(a). Therefore, ${\cal K}( \tilde{\cal P}, [0, \tau_{\mycyc})) \leq (1 + \eps) \cdot {\cal K}( {\cal P}^*,  [ \tau_{\kappa}, \tau_{\kappa+1}))$.
\end{itemize}
By aggregating these two observations, we have
\begin{equation} \label{eqn:step2_feas_obj}
C( \tilde{\cal P} ) ~~=~~ \frac{  C( \tilde{\cal P}, [0,\tau_{\mycyc})) }{ \tau_\mycyc } ~~\leq~~ (1 + \eps) \cdot \frac{ C( {\cal P^*}, [ \tau_{\kappa}, \tau_{\kappa+1}) ) }{ \tau_{\kappa+1} - \tau_{\kappa} } ~~\leq~~ (1 + 3\eps) \cdot C( {\cal P^*} ) \ ,
\end{equation}
where the last inequality holds since $\frac{ C( {\cal P^*}, [ \tau_{\kappa}, \tau_{\kappa+1}) ) }{ \tau_{\kappa+1} - \tau_{\kappa} } \leq (1 + \eps) \cdot C( {\cal P^*} )$ according to~\eqref{eqn:proof_lem_good_cyclic_eq1}.

\paragraph{Proof of item~2.} We first establish an auxiliary claim about how the optimal long-run average cost $C( {\cal P^*})$ is related to ${\cal M}$. The proof of this result, whose details are provided in  Appendix~\ref{app:proof_clm_Cpstar_vs_M}, compares the optimal dynamic policy ${\cal P}^*$ to a benchmark that independently replenishes each commodity via an optimal SOSI policy,  subject to our warehouse capacity constraint. Then, basic convexity properties yield an aggregate upper bound on $C( {\cal P}^* )$, captured by the definition of ${\cal M}$.

\begin{claim} \label{clm:Cpstar_vs_M}
$C( {\cal P}^* ) \leq n {\cal M}$.
\end{claim}

We proceed by explaining why the above claim leads to the desired bounds on the cycle length $\tau_{\mycyc}$, specifically showing that $\tau_{\mycyc} \in [\frac{ K_{\max} }{ 2\eps n {\cal M} }, \frac{ 2 n {\cal M} }{ \eps^2 H_{\min} }]$:
\begin{itemize}
    \item {\em Upper bound}: By definition of $\tau_{\kappa+1}$, we know that there exists at least one commodity $i \in [n]$ with $N( \tilde{\cal P}_i, [0,\tau_{\mycyc})) = \frac{ 1 }{ \eps }$ orders. At least one of these orders consists of at least $\eps \tau_{\mycyc}$ units, meaning that the holding cost of these units is at least $H_i \eps^2 \tau_{\mycyc}^2$. As a result,
    \begin{equation} \label{eqn:proof_good_cyclic_item2}
    H_i \eps^2 \tau_{\mycyc}^2 ~~\leq~~ C( \tilde{\cal P}, [0,\tau_{\mycyc}] ) ~~\leq~~ (1 + 3\eps) \cdot \tau_{\mycyc} \cdot  C( {\cal P^*} ) ~~\leq~~ 2 \tau_{\mycyc} n {\cal M} \ ,
    \end{equation}
    where the second and third inequalities hold since $\frac{  C( \tilde{\cal P}, [0,\tau_{\mycyc})) }{ \tau_\mycyc } \leq (1 + 3\eps) \cdot C( {\cal P^*} )$, as shown  in~\eqref{eqn:step2_feas_obj}, and since $C( {\cal P^*} ) \leq n {\cal M}$ according to Claim~\ref{clm:Cpstar_vs_M}. By rearranging, we indeed get $\tau_{\mycyc} \leq \frac{ 2 n {\cal M} }{ \eps^2 H_{i} } \leq \frac{ 2 n {\cal M} }{ \eps^2 H_{\min} }$.

    \item {\em Lower bound}: Again by definition of $\tau_{\kappa+1}$, we know that every commodity $i \in [n]$ has $N( \tilde{\cal P}_i, [0,\tau_{\mycyc })) \geq \frac{ 1 }{ \eps }$ orders in each cycle. Therefore, $\tilde{\cal P}$ incurs an ordering cost of at least $\frac{ K_{\max} }{ \eps }$ in each cycle,  implying that
    \[ \frac{ K_{\max} }{ \eps \tau_{\mycyc } } ~~\leq~~ C( \tilde{\cal P} ) ~~\leq~~ (1 + 3\eps) \cdot C( {\cal P^*} ) ~~\leq~~ 2 n {\cal M} \ . \]
    Here, the second inequality is precisely~\eqref{eqn:step2_feas_obj}, and the third inequality follows from Claim~\ref{clm:Cpstar_vs_M}. Again by rearranging, $\tau_{\mycyc } \geq \frac{ K_{\max} }{ 2\eps n {\cal M} }$.
\end{itemize}

\paragraph{Proof of item~3.} As explained while defining $\tilde{\cal P}$, we know that for every commodity $i \in [n]$, this policy places $N( \tilde{\cal P}_i, [0,\tau_\mycyc)) \geq \frac{ 1 }{ \eps }$ orders across the segment $[0,\tau_\mycyc)$, with equality for at least one commodity. In the opposite direction, we obtain an upper bound on $N( \tilde{\cal P}_i, [0,\tau_\mycyc))$ by observing that, since the ordering cost of each commodity $i$ by itself is $K_i \cdot N( \tilde{\cal P}_i, [0,\tau_\mycyc))$, it follows that
\[ K_i \cdot N( \tilde{\cal P}_i, [0,\tau_\mycyc)) ~~\leq~~ C( \tilde{\cal P}, [0,\tau_\mycyc) ) ~~\leq~~ 2 \tau_\mycyc n {\cal M} ~~\leq~~ \frac{ 4 n^2 {\cal M}^2 }{ \eps^2 H_{\min} } \ . \]
Here, the second inequality has already been derived in~\eqref{eqn:proof_good_cyclic_item2}, starting from the second term, and the third inequality holds since $\tau_{\mycyc} \leq \frac{ 2 n {\cal M} }{ \eps^2 H_{\min} }$, as shown in item~2. By rearranging this inequality, we get $N( \tilde{\cal P}_i, [0,\tau_\mycyc)) \leq \frac{ 4 n^2 {\cal M}^2 }{ \eps^2 K_i H_{\min} } \leq \frac{ 4 n^2 {\cal M}^2 }{ \eps^2 K_{\min} H_{\min} }$.

\subsection{Milestone~II: Frequency classes and the Alignment Theorem} \label{subsec:structure_thm}

\paragraph{Frequency classes.} It is important to emphasize that, while Theorem~\ref{thm:good_cyclic}  provides a proof of existence,  the near-optimal cyclic policy $\tilde{\cal P}$ is obviously unknown from an algorithmic perspective. Toward arriving at a constructive dynamic programming approach in Section~\ref{sec:approx_scheme_details}, we wish to instill further useful structure, which is not necessarily satisfied by $\tilde{\cal P}$. For this purpose, let us focus on a single cycle $[0, \tau_\mycyc)$ of this policy. By Theorem~\ref{thm:good_cyclic}(3), the latter segment has at most ${\cal U} = \frac{ 4 n^2 {\cal M}^2 }{ \eps^2 K_{\min} H_{\min} }$ orders of each commodity, and we can therefore partition the set of commodities into frequency classes $\tilde{\cal F}_1, \ldots, \tilde{\cal F}_Q$ as follows:
\begin{itemize}
    \item The class $\tilde{\cal F}_1$ consists of commodities with $N( \tilde{\cal P}_i, [0,\tau_{\mycyc})) \in [1,(\frac{n}{\eps})^{3}]$ orders.

    \item The class $\tilde{\cal F}_2$ is comprised of those with $N( \tilde{\cal P}_i, [0,\tau_{\mycyc})) \in ((\frac{n}{\eps})^{3}, (\frac{n}{\eps})^{6}]$ orders.

    \item We continue this construction up to $\tilde{\cal F}_Q$, consisting of commodities with $N( \tilde{\cal P}_i, [0,\tau_{\mycyc})) \in ((\frac{n}{\eps})^{3(Q-1)}, (\frac{n}{\eps})^{3Q}]$. Here, we set $Q = \lceil \log_{(n/\eps)^{3}} {\cal U}  \rceil$, thereby ensuring that the union of $\tilde{\cal F}_1, \ldots, \tilde{\cal F}_Q$  contains all commodities.
\end{itemize}

\begin{figure}[htbp!]
    \centering
    \resizebox{\textwidth}{!}{%
    \begin{tikzpicture}[x=10cm, y=1cm]
        \def\taucycle{1}
        \def\nCoarse{8}
        \def\nFine{40}

        \draw[thick, ->] (-0.05,0) -- (\taucycle + 0.05, 0) node[right] {$t$};

        \foreach \i in {0, ..., \nFine} {
            \draw[blue!50, thick] (\i/\nFine, -0.15) -- (\i/\nFine, 0.15);
        }

        \foreach \i in {0, ..., \nCoarse} {
            \draw[red!80, line width=2pt] (\i/\nCoarse, -0.3) -- (\i/\nCoarse, 0.3);
        }

        \draw[black!60!green, line width=1.2pt]
            (0,0) -- (0, 0.4) -- (0.05, 0) -- (0.05, 0.6) -- (0.125, 0)
            -- (0.125, 1.0) -- (0.25, 0)
            -- (0.25, 0.2) -- (0.275, 0) -- (0.275, 0.2) -- (0.300, 0)
            -- (0.300, 0.2) -- (0.325, 0) -- (0.325, 0.4) -- (0.375, 0)
            -- (0.375, 0.6) -- (0.45, 0) -- (0.45, 0.4) -- (0.5, 0)
            -- (0.5, 1.0) -- (0.625, 0)
            -- (0.625, 0.4) -- (0.675, 0) -- (0.675, 0.6) -- (0.75, 0)
            -- (0.75, 1.0) -- (0.875, 0)
            -- (0.875, 0.4) -- (0.925, 0) -- (0.925, 0.6) -- (1.0, 0);

        \node[black!60!green, anchor=south, font=\small\bfseries\boldmath] at (0.45, 1.25) {$B$-aligned policy for commodity $i \in \tilde{\cal F}_q$};

        \node[below=0.25cm] at (0,0) {0};
        \node[below=0.25cm] at (\taucycle,0) {$\tau_{\mathrm{cycle}}$};

        \node[blue!80, font=\scriptsize, anchor=north] (labelFine) at (0.05, -0.6) {orders allowed ($B_q^+$)};
        \draw[blue!50, thin, ->] (labelFine.north) -- (0.05, -0.15);

        \node[red!80, font=\scriptsize, anchor=north] (labelCoarse) at (0.375, -0.6) {zero inventory ($B_q^-$)};
        \draw[red!80, thin, ->] (labelCoarse.north) -- (0.375, -0.3);

        \begin{scope}[shift={(\taucycle, 2.2)}, scale=0.7]
            \begin{scope}[x=1cm, y=1cm]
                \draw[black!60!green, line width=1.2pt] (-2.0, 0.2) -- (-1.5, 0.2);
                \node[anchor=west, font=\footnotesize, inner sep=0pt, xshift=2pt] at (-1.5, 0.2) {inventory};

                \draw[blue!50, thick] (-2.0, -0.3) -- (-1.5, -0.3);
                \node[anchor=west, font=\footnotesize, inner sep=0pt, xshift=2pt] at (-1.5, -0.3) {$B_q^+$ (fine)};

                \draw[red!80, line width=2pt] (-2.0, -0.8) -- (-1.5, -0.8);
                \node[anchor=west, font=\footnotesize, inner sep=0pt, xshift=2pt] at (-1.5, -0.8) {$B_q^-$ (coarse)};
            \end{scope}
        \end{scope}

    \end{tikzpicture}%
    }
    \caption{Visual representation of a {$B$}-aligned policy with respect to our nested breakpoint structure. The inventory curve reaches zero at every coarse breakpoint ({$B_q^-$}), placing orders only at fine breakpoints ({$B_q^+$}).}
    \label{fig:baligned_policy}
\end{figure}
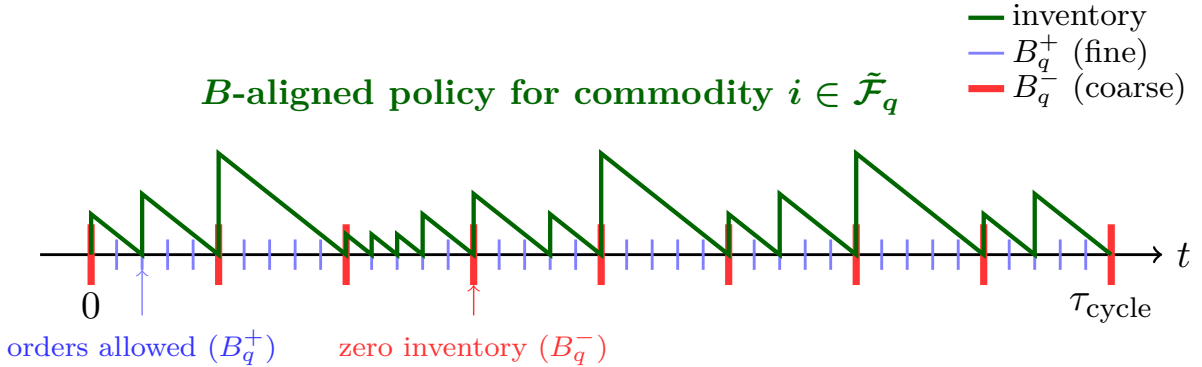

\paragraph{Subdivisions and the Alignment Theorem.} For every $q \in [Q]$, suppose we subdivide the segment $[0,\tau_{\mycyc})$ into
$(\frac{n}{\eps})^{3q + 1}$ equal-length subintervals; the collection of breakpoints obtained will be denoted by $B_q^+$. Concurrently, we define a more coarse subdivision of $[0,\tau_{\mycyc})$, partitioning this segment into $(\frac{n}{\eps})^{3q-4}$ equal-length subintervals. Here, $B_q^-$ will designate the set of resulting breakpoints, noting that $B_q^- \subseteq B_q^+$. With respect to these subdivisions, we will be limiting our attention to a structured family of policies, referred to as being $B$-aligned, schematically illustrated in Figure~\ref{fig:baligned_policy}. To this end, we say that a zero-inventory ordering policy ${\cal P}$ for the segment $[0,\tau_{\mycyc})$ is $B$-aligned when, for every frequency class $q \in [Q]$  and commodity $i \in \tilde{\cal F}_q$, the next two properties are simultaneously satisfied:
\begin{enumerate}
    \item \label{item:B_aligned_1} {\em Zero inventory at $B_q^-$-points:} $I( {\cal P}_i, b^-) = \lim_{t \uparrow b} I( {\cal P}_i, t) = 0$, for every $b \in B_q^-$.

    \item \label{item:B_aligned_2} {\em Orders only at $B_q^+$-points:} ${\cal P}_i$ places orders only at points in $B_q^+$.
\end{enumerate}

Our main structural result in this context, called the Alignment Theorem, proves the existence of $B$-aligned policies that are near-feasible and near-optimal at the same time. The specifics of this finding are summarized in Theorem~\ref{thm:exist_aligned}, whose proof is provided in  Section~\ref{subsec:proof_lem_exist_aligned}. While the finer details of our proof are subtle, at least concentually, property~\ref{item:B_aligned_1} will be attained by initially placing new orders at all $B_q^-$-points, whereas property~\ref{item:B_aligned_2} will subsequently require stretching and possibly discarding certain orders. Combined with Theorem~\ref{thm:good_cyclic}, we will argue that these alterations have minor consequences in terms of space and cost.

\begin{theorem}[Alignment]\label{thm:exist_aligned}
There exists a $B$-aligned policy $\hat{\cal P}$ satisfying the next two properties:
\begin{enumerate}
    \item {\em Space}: $V_{\max}( \hat{\cal P} ) \leq (1+\eps) \cdot {\cal V}$.

    \item {\em Cost}: $C( \hat{\cal P}, [0,\tau_{\mycyc}) ) \leq (1 + \eps) \cdot C( \tilde{\cal P}, [0,\tau_{\mycyc}))$.
\end{enumerate}
\end{theorem}

\subsection{Proof of Theorem~\ref{thm:exist_aligned}} \label{subsec:proof_lem_exist_aligned}

To establish the existence of a $B$-aligned policy $\hat{\cal P}$ which is near-feasible and near-optimal, we alter $\tilde{\cal P}$ in two  steps. As explained below, the first step is meant to attain property~\ref{item:B_aligned_1}, whereas the second step will ensure property~\ref{item:B_aligned_2}.

\paragraph{Ensuring property~\ref{item:B_aligned_1}: Zero inventory at {\boldmath ${B_q^-}$}-points.} For every frequency class $q \in [Q]$ and commodity $i \in \tilde{\cal F}_q$, we augment $\tilde{\cal P}$ with an additional $i$-order at every $B_q^-$-point. It is easy to verify that the ordering quantity of each $i$-order can be adjusted such that the resulting policy $\hat{\cal P}^1$ satisfies the next two properties:
\begin{itemize}
    \item We arrive to $B_q^-$-points with zero inventory, i.e., $I( \hat{\cal P}^1_i, b^-) = 0$ for every $b \in B_q^-$, thereby instilling property~\ref{item:B_aligned_1}.

    \item The inventory level of this commodity across the entire cycle is upper-bounded by its analogous level with respect to $\tilde{\cal P}$, namely, $I( \hat{\cal P}^1_i, t) \leq I( \tilde{\cal P}_i, t)$ for every $t \in [0, \tau_\mycyc)$.
\end{itemize}
As a consequence, $\hat{\cal P}^1$ is capacity-feasible, and its holding cost does not exceed that of $\tilde{\cal P}$. In terms of ordering costs, each commodity $i \in \tilde{\cal F}_q$ previously had $N( \tilde{\cal P}_i, [0,\tau_\mycyc)) \geq (\frac{n}{\eps})^{3(q-1)}$ orders, by definition of $\tilde{\cal F}_q$, whereas the number of newly-added $i$-orders is at most
\[ |B_q^-| ~~=~~ \left( \frac{n}{\eps} \right)^{3q-4}   ~~\leq~~ \frac{ \eps }{ n } \cdot N( \tilde{\cal P}_i, [0,\tau_\mycyc)) \ . \]
Therefore, the combined ordering cost of $\hat{\cal P}^1$ across the entire cycle is
\begin{align}
{\cal K}( \hat{\cal P}^1, [0,\tau_\mycyc) ) & ~~=~~ \sum_{q \in [Q]} \sum_{i \in \tilde{\cal F}_q} K_i \cdot N( \hat{\cal P}^1_i, [0,\tau_\mycyc) ) \nonumber \\
& ~~\leq~~ \sum_{q \in [Q]} \sum_{i \in \tilde{\cal F}_q} K_i \cdot \left( N( \tilde{\cal P}_i, [0,\tau_\mycyc) ) + |B_q^-|\right) \nonumber\\
& ~~\leq~~ \left( 1 + \frac{ \eps }{ n } \right) \cdot \sum_{q \in [Q]} \sum_{i \in \tilde{\cal F}_q} K_i \cdot N( \tilde{\cal P}_i, [0,\tau_\mycyc)) \nonumber \\
& ~~\leq~~ (1 + \eps) \cdot {\cal K}( \tilde{\cal P}, [0,\tau_\mycyc) ) \ . \label{eqn:proof_thm_aligned_A}
\end{align}

\paragraph{Ensuring property~\ref{item:B_aligned_2}: Orders only at {\boldmath ${B_q^+}$}-points.} For every frequency class $q \in [Q]$ and commodity $i \in \tilde{\cal F}_q$, the remaining issue with $\hat{\cal P}^1$ is that this policy may be placing $i$-orders in-between points in $B_q^+$. To eliminate this issue, we modify each $i$-order as follows:
\begin{itemize}
    \item Let $[t,t+\Delta)$ be the interval covered by the currently considered $i$-order. Specifically, since $\hat{\cal P}^1_i$ is a zero-inventory ordering policy, it reaches  time $t$ with zero inventory, ordering $\Delta$ units at that time.

    \item {\em Right-lengthening}: We extend the covered interval further to the right, by setting its right endpoint as $\lceil t + \Delta \rceil^{ B_q^+ }$. Here, $\lceil \cdot \rceil^{ B_q^+ }$ is an operator that rounds its argument up to the next $B_q^+$-point. This step requires increasing our ordering quantity from $\Delta$ to $\lceil t + \Delta \rceil^{ B_q^+ } - t \leq \Delta + \frac{ \tau_\mycyc }{ (n/\eps)^{3q + 1} }$, where the last inequality holds since the gap between successive $B_q^+$-points is $\frac{ \tau_\mycyc }{ (n/\eps)^{3q + 1} }$.

    \item {\em Left-shortening}: We now shorten the newly-covered interval, $[t,\lceil t + \Delta \rceil^{ B_q^+ })$, by setting its left endpoint as $\lceil t \rceil^{ B_q^+ }$, which requires decreasing our ordering quantity. In the degenerate event where $[t,t+\Delta)$ shrinks into a single point, i.e., $\lceil t \rceil^{ B_q^+ } = \lceil t + \Delta \rceil^{ B_q^+ }$, we discard this order.
\end{itemize}
It is easy to verify that the resulting policy $\hat{\cal P}^2$ covers the entire cycle $[0,\tau_\mycyc)$, places orders only at $B_q^+$-points, and preserves property~\ref{item:B_aligned_1}. We proceed by analyzing this policy in terms of its occupied space and cost, respectively corresponding to Claims~\ref{clm:hatP2_near_feasible} and~\ref{clm:hatP2_near_optimal} below.

\begin{claim} \label{clm:hatP2_near_feasible}
$\hat{\cal P}^2$ is $(1+\eps) $-feasible.
\end{claim}
\begin{proof}
As explained above, for every frequency class $q \in [Q]$ and commodity $i \in \tilde{\cal F}_q$, right-lengthening $i$-orders may increase the inventory level of this commodity at any point by at most $\frac{ \tau_\mycyc }{ (n/\eps)^{3q + 1} }$, whereas left-shortening may only decrease this level. Therefore, for every $t \in [0,\tau_\mycyc)$,
\begin{equation} \label{eqn:proof_clm_hatP2_near_feasible_eq1}
I( \hat{\cal P}^2_i, t) ~~\leq~~ I( \hat{\cal P}^1_i, t) + \frac{ \tau_\mycyc }{ (n/\eps)^{3q + 1} } \ .
\end{equation}
Based on this observation, we conclude that the peak space occupied by $\hat{\cal P}^2$ is at most $(1 + \eps) \cdot {\cal V}$, since
\begin{align}
V_{\max}( \hat{\cal P}^2 ) & ~~=~~ \max_{t \in [0,\tau_\mycyc)}  V( \hat{\cal P}^2, t) \nonumber \\
& ~~=~~ \max_{t \in [0,\tau_\mycyc)}  \sum_{q \in [Q]} \sum_{i \in \tilde{\cal F}_q} \gamma_i \cdot I( \hat{\cal P}^2_i, t) \nonumber \\
& ~~\leq~~ \max_{t \in [0,\tau_\mycyc)}  \sum_{q \in [Q]} \sum_{i \in \tilde{\cal F}_q} \gamma_i \cdot \left( I( \hat{\cal P}^1_i, t) + \frac{ \tau_\mycyc }{ (n/\eps)^{3q + 1} } \right) \nonumber \\
& ~~\leq~~ V_{\max}( \hat{\cal P}^1 ) + \eps {\cal V} \label{eqn:proof_thm_aligned_1} \\
& ~~\leq~~ V_{\max}( \tilde{\cal P} ) + \eps {\cal V} \label{eqn:proof_thm_aligned_2} \\
& ~~\leq~~ (1 + \eps) \cdot {\cal V} \ . \label{eqn:proof_thm_aligned_3}
\end{align}
Here, inequality~\eqref{eqn:proof_thm_aligned_1} is obtained by arguing that $\gamma_i \cdot \frac{ \tau_{\mycyc} }{ (n/\eps)^{3q} } \leq {\cal V}$ for every commodity $i \in \tilde{\cal F}_q$. Indeed, we have $N( \tilde{\cal P}_i, [0,\tau_{\mycyc})) \leq (\frac{n}{\eps})^{3q}$ by definition of $\tilde{\cal F}_q$, implying that our original policy $\tilde{\cal P}$ has at least one $i$-order consisting of at least $\frac{ \tau_{\mycyc} }{ (n/\eps)^{3q} }$ units. Since $\tilde{\cal P}$ is capacity-feasible, we must have $\gamma_i \cdot \frac{ \tau_{\mycyc} }{ (n/\eps)^{3q} } \leq {\cal V}$. Inequality~\eqref{eqn:proof_thm_aligned_2} holds since $I( \hat{\cal P}^1_i, t) \leq I( \tilde{\cal P}_i, t)$ for every $t \in [0, \tau_\mycyc)$, as explained when establishing property~\ref{item:B_aligned_1}. Finally, inequality~\eqref{eqn:proof_thm_aligned_3} follows by recalling that the policy  $\tilde{\cal P}$ is capacity-feasible, according to Theorem~\ref{thm:good_cyclic}.
\end{proof}

\begin{claim} \label{clm:hatP2_near_optimal}
$C( \hat{\cal P}^2, [0,\tau_{\mycyc}) ) \leq (1 + \eps) \cdot C( \tilde{\cal P}, [0,\tau_{\mycyc}))$.
\end{claim}
\begin{proof}
First, we can relate the total ordering cost of $\hat{\cal P}^2$ across $[0,\tau_{\mycyc})$ to the analogous cost with respect to $\tilde{\cal P}$ by observing that
\[ {\cal K}( \hat{\cal P}^2, [0,\tau_{\mycyc}) ) ~~\leq~~ {\cal K}( \hat{\cal P}^1, [0,\tau_{\mycyc}) ) ~~\leq~~ ( 1 + \eps) \cdot {\cal K}( \tilde{\cal P}, [0,\tau_{\mycyc})) \ . \]
Here, the first inequality holds since our method for converting  $\hat{\cal P}^1$ into $\hat{\cal P}^2$ does not involve placing additional orders; in fact, some orders may be discarded during left-shortening. The second inequality is precisely~\eqref{eqn:proof_thm_aligned_A}.

Moving to consider holding costs, let us recall that inequality~\eqref{eqn:proof_clm_hatP2_near_feasible_eq1} compares between the inventory levels of $\hat{\cal P}^2$ and $\hat{\cal P}^1$, stating that $I( \hat{\cal P}^2_i, t) \leq I( \hat{\cal P}^1_i, t) + \frac{ \tau_\mycyc }{ (n/\eps)^{3q + 1} }$ for every frequency class $q \in [Q]$, commodity $i \in \tilde{\cal F}_q$, and time point $t \in [0,\tau_\mycyc)$. As a result,
\begin{align}
{\cal H}( \hat{\cal P}^2, [0,\tau_{\mycyc}) ) & ~~\leq~~ \sum_{q \in [Q]} \sum_{i \in \tilde{\cal F}_q} \left( {\cal H}( \hat{\cal P}^1_i, [0,\tau_{\mycyc}) )  + 2H_i \cdot \frac{ \tau_\mycyc^2 }{ (n/\eps)^{3q + 1} } \right) \nonumber \\
& ~~=~~ {\cal H}( \hat{\cal P}^1, [0,\tau_{\mycyc}) ) +2 \cdot \sum_{q \in [Q]} \sum_{i \in \tilde{\cal F}_q} H_i \cdot \frac{ \tau_\mycyc^2 }{ (n/\eps)^{3q + 1} } \nonumber  \\
& ~~\leq~~ {\cal H}( \tilde{\cal P}, [0,\tau_{\mycyc}) ) + 2 \cdot \sum_{q \in [Q]} \sum_{i \in \tilde{\cal F}_q} H_i \cdot \frac{ \tau_\mycyc^2 }{ (n/\eps)^{3q + 1} } \label{eqn:proof_thm_aligned_4} \\
& ~~\leq~~ {\cal H}( \tilde{\cal P}, [0,\tau_{\mycyc}) ) + \frac{ 2\eps }{ n } \cdot \sum_{q \in [Q]} \sum_{i \in \tilde{\cal F}_q} {\cal H}( \tilde{\cal P}_i, [0,\tau_{\mycyc}) ) \label{eqn:proof_thm_aligned_5} \\
& ~~\leq~~ (1 + \eps) \cdot {\cal H}( \tilde{\cal P}, [0,\tau_{\mycyc}) ) \ . \nonumber
\end{align}
Here, inequality~\eqref{eqn:proof_thm_aligned_4} holds since, as explained when establishing property~\ref{item:B_aligned_1}, the holding cost of $\hat{\cal P}^1$ is upper-bounded by that of $\tilde{\cal P}$. Inequality~\eqref{eqn:proof_thm_aligned_5} follows from Claim~\ref{clm:aux_holding_inequality} below, whose proof is provided in Appendix~\ref{app:proof_clm_aux_holding_inequality}. Its high-level intuition is that, according to our definition of frequency classes, we know that $\tilde{\cal P}$ places at most $(\frac{n}{\eps})^{3q}$ orders for each commodity $i \in \tilde{\cal F}_q$. In turn, the overall holding cost ${\cal H}( \tilde{\cal P}_i, [0,\tau_{\mycyc}) )$ of this commodity can be lower-bounded via the one incurred by optimizing the placement of these orders, thereby producing the next auxiliary claim.

\begin{claim} \label{clm:aux_holding_inequality}
$H_i \cdot \frac{ \tau_\mycyc^2 }{ (n/\eps)^{3q} } \leq {\cal H}( \tilde{\cal P}_i, [0,\tau_{\mycyc}) )$, for every $q \in [Q]$ and $i \in \tilde{\cal F}_q$.
\end{claim}
\end{proof} 
\section{The Approximation Scheme}
\label{sec:approx_scheme_details}

In this section, we exploit the Alignment Theorem to devise an approximate dynamic programming approach for computing near-optimal capacity-feasible replenishment policies. Specifically, as stated in Theorem~\ref{thm:PTAS_exponential}, given any accuracy level $\eps > 0$, our $O( | {\cal I} |^{O(n)} \cdot 2^{ O( n^{6} / \eps^{5} ) } )$-time algorithm constructs an efficiently-representable cyclic policy whose long-run average cost is within factor $1 + \eps$ of optimal.

\paragraph{Outline.} Moving forward, Section~\ref{subsec:recusive_main_result} articulates the approximation guarantee we are aiming for and identifies the optimal substructure implied by $B$-aligned policies. Conceptually, this characterization allows us to process the frequency classes $\tilde{\cal F}_1, \ldots, \tilde{\cal F}_Q$ one after the other, by defining a novel interface between low- and high-frequency commodities at any point in time. To efficiently capture the latter interface, Section~\ref{subsec:approx_stat} introduces a condensed state description, with an exact inventory representation of our actions in the preceding recursion level, coupled with a proxy representation of our aggregate actions in all earlier levels. Sections~\ref{subsec:dp_guess_State} and~\ref{subsec:recursive_details} elaborate on the finer details of our approximate dynamic program, showing that $B$-aligned policies lead to an ordering decomposition across successive breakpoint intervals. Finally, Section~\ref{subsec:DP_analysis} completes the proof of Theorem~\ref{thm:PTAS_exponential} by arguing that this program indeed produces a near-optimal capacity-feasible replenishment policy, while still admitting an $O( | {\cal I} |^{O(n)} \cdot 2^{ O( n^{6} / \eps^{5} ) } )$-time implementation.

\subsection{Main result and optimal substructure} \label{subsec:recusive_main_result}

\paragraph{Intent.} For the purpose of computing a near-optimal capacity-feasible policy, letting $\tilde{\cal P}$ and $\hat{\cal P}$ be the policies whose existence has been established, respectively, in Theorems~\ref{thm:good_cyclic} and~\ref{thm:exist_aligned}, we know that
\[ C( \hat{\cal P} ) ~~\leq~~ (1 + \eps) \cdot C( \tilde{\cal P}) ~~\leq~~ (1 + 7\eps) \cdot \opt\eqref{eqn:model_warehouse} \ . \]
In addition, $\hat{\cal P}$ is $B$-aligned and $(1+\eps)$-feasible, in the sense that $V_{\max}( \hat{\cal P} ) \leq (1+\eps) \cdot {\cal V}$. Therefore, in terms of space requirement and cost, it suffices to obtain a $(1 + O(\eps))$-feasible policy ${\cal P}$ for the interval $[0, \tau_\mycyc)$ with $C( {\cal P}, [0,\tau_{\mycyc}) ) = (1+O(\eps)) \cdot C( \hat{\cal P}, [0,\tau_{\mycyc}) )$. When this policy is scaled by a factor of $1 + O(\eps)$, we end up with a capacity-feasible policy whose long-run average cost is within factor $1 + O(\eps)$ of optimal. In terms of efficiency, we will show that our algorithm can be implemented in $O( | {\cal I} |^{O(n)} \cdot 2^{ O( n^{6} / \eps^{5} ) } )$ time.

\paragraph{Notation.} For the remainder of this section, let  ${\cal P}^{B*}$ be an optimal $B$-aligned $(1+\eps)$-feasible policy for the segment $[0,\tau_{\mycyc})$, noting that $C( {\cal P}^{B*}, [0,\tau_{\mycyc}) ) \leq C( \hat{\cal P}, [0,\tau_{\mycyc}))$. We remind the reader that, for any $q \in [Q]$, the segment $[0, \tau_\mycyc)$ was subdivided into $(\frac{n}{\eps})^{3q-4}$ equal-length subintervals, thereby creating the set of breakpoints $B_q^-$. By property~\ref{item:B_aligned_1} of $B$-aligned policies, we know that for every commodity $i \in \tilde{\cal F}_q$, the policy ${\cal P}^{B*}$ reaches zero inventory at all $B_q^-$-points. Moreover, since $B_1^- \subseteq \cdots \subseteq B_Q^-$, this policy actually  reaches zero inventory at $B_q^-$-points for every commodity belonging to one of the classes $\tilde{\cal F}_q, \ldots, \tilde{\cal F}_Q$. We use $\tilde{\cal F}_{\geq q}$ to designate their union, with $\tilde{\cal F}_{\leq q-1}$ being the union of $\tilde{\cal F}_1, \ldots, \tilde{\cal F}_{q-1}$.

\paragraph{Optimal substructure.} To unveil the optimal substructure that underlies our dynamic program, consider the following question:
\begin{quote}
{\em For any $q \in [Q]$, what is the dependency between the actions taken by ${\cal P}^{B*}$ for $\tilde{\cal F}_{\geq q}$-commodities and its actions for $\tilde{\cal F}_{\leq q-1}$-commodities?}
\end{quote}
To address this question, let us examine an interval $[b_\mathrm{entry}, b_\mathrm{exit}]$ that stretches between two successive points in $B_q^-$. Since ${\cal P}^{B*}$ is $B$-aligned, it follows that $I({\cal P}^{B*}_i, b_\mathrm{entry}^-) = I({\cal P}^{B*}_i, b_\mathrm{exit}^-) = 0$ for every commodity $i \in \tilde{\cal F}_{\geq q}$. Thus, we can assume without loss of generality that the actions taken by ${\cal P}^{B*}$ for $\tilde{\cal F}_{\geq q}$-commodities across $[b_\mathrm{entry}, b_\mathrm{exit}]$ only depend on two types of statistics, summarizing its actions for $\tilde{\cal F}_{\leq q-1}$-commodities:
\begin{itemize}
    \item {\em $\tilde{\cal F}_{q-1}$-inventory levels}: By property~\ref{item:B_aligned_2} of $B$-aligned policies, it follows that ${\cal P}^{B*}$ places orders for every commodity $i \in \tilde{\cal F}_{q-1}$ only at $B_{q-1}^+$-points. Since $|B_q^-| = (\frac{n}{\eps})^{3q-4}$ and $|B_{q-1}^+| = (\frac{n}{\eps})^{3(q-1)+1}$, we know that the number of $B_{q-1}^+$-points in $[b_\mathrm{entry}, b_\mathrm{exit}]$ is exactly $(\frac{n}{\eps})^2$. Once the inventory levels of commodity $i$ at these points (including $b_\mathrm{entry}$ and $b_\mathrm{exit}$) are known, they uniquely determine its inventory level across the entire interval $[b_\mathrm{entry}, b_\mathrm{exit}]$.

    \item {\em $\tilde{\cal F}_{\leq q-2}$-entry-inventory levels}: By the same argument, for every commodity $i \in \tilde{\cal F}_{\leq q-2}$, any $i$-order placed by ${\cal P}^{B*}$ resides in $B_{q-2}^+$. Since $B_{q-2}^+ \subseteq B_q^-$, it follows that the interval $[b_\mathrm{entry}, b_\mathrm{exit}]$ does not have any $i$-order in its interior. As a result, for such commodities, the optimal policy ${\cal P}^{B*}$ only has to account for the inventory level $I({\cal P}^{B*}_i, b_\mathrm{entry}^+) = \lim_{t \downarrow b_\mathrm{entry}} I({\cal P}^{B*}_i,t)$  immediately after entering this interval.
\end{itemize}
In fact, once we know the above-mentioned statistics, the interval $[b_\mathrm{entry}, b_\mathrm{exit}]$ becomes superfluous, in the sense that ${\cal P}^{B*}$ would act in an identical way between any pair of successive $B_q^-$-points  with these statistics.

\subsection{Approximate statistics} \label{subsec:approx_stat}

\paragraph{Exact representation for {\boldmath ${\tilde{\cal F}_{q-1}}$}-inventory levels.} In terms of their representation within our dynamic program, $\tilde{\cal F}_{q-1}$-inventory levels will be relatively easy to keep track of. As mentioned above, for any commodity $i \in \tilde{\cal F}_{q-1}$, there are $(\frac{n}{\eps})^2$ points in $B_{q-1}^+$ where $i$-orders can be placed along $[b_\mathrm{entry}, b_\mathrm{exit}]$. In addition, it is not difficult to verify that the latest $i$-order to the left of $b_\mathrm{entry}$ is restricted to being one of
$(\frac{n}{\eps})^{5}$ points in $B_{q-1}^+$. Indeed, by property~\ref{item:B_aligned_1} of $B$-aligned policies, we know that this commodity reaches zero inventory at $B_{q-1}^-$-points. Therefore, there must be at least one $i$-order between the latest $B_{q-1}^-$-point to the left of $b_\mathrm{entry}$ and $b_\mathrm{entry}$ itself. This  interval has at most $\frac{ | B_{q-1}^+ | }{ | B_{q-1}^- | } = (\frac{n}{\eps})^{5}$ points in $B_{q-1}^+$, which are the only ones where $i$-orders can be placed, by property~\ref{item:B_aligned_2}. Via a symmetric argument, precisely the same claim applies to the earliest $i$-order to the right of $b_\mathrm{exit}$. Once we determine the orders placed at each of these points, for which there are $( 2^{ (n/\eps)^2 } \cdot (\frac{n}{\eps})^{10} )^{ |\tilde{\cal F}_{q-1}| } = O(2^{ O( n^3 / \eps^2 ) })$ options over all $\tilde{\cal F}_{q-1}$-commodities, they uniquely determine $\tilde{\cal F}_{q-1}$-inventory levels.

\paragraph{Proxy representation for {\boldmath ${\tilde{\cal F}_{\leq q-2}}$}-entry-inventory levels.} In contrast, regarding the number of configurations taken by the $\tilde{\cal F}_{\leq q-2}$-entry-inventory levels $\{ I({\cal P}^{B*}_i, b_\mathrm{entry}^+) \}_{i \in \tilde{\cal F}_{\leq q-2}}$, it is unclear how this quantity can be upper-bounded to even remotely meet the intended running time of Theorem~\ref{thm:PTAS_exponential}. For this reason, we propose a one-dimensional statistic that will serve as a sufficiently-good alternative. To understand the motivation for this statistic, note that since the interval $[b_\mathrm{entry}, b_\mathrm{exit}]$ does not have  $\tilde{\cal F}_{\leq q-2}$-orders in its interior, the inventory level of each commodity $i \in \tilde{\cal F}_{\leq q-2}$ linearly drops by exactly $b_\mathrm{exit} - b_\mathrm{entry}$ units. As such, the total space requirement $\sum_{i \in \tilde{\cal F}_{\leq q-2}} \gamma_i \cdot I({\cal P}^{B*}_i, b_\mathrm{exit}^-)$ of the $\tilde{\cal F}_{\leq q-2}$-commodities just before exiting $[b_\mathrm{entry}, b_\mathrm{exit}]$ is related to its analogous quantity $\sum_{i \in \tilde{\cal F}_{\leq q-2}} \gamma_i \cdot  I({\cal P}^{B*}_i, b_\mathrm{entry}^+)$ immediately after entering this interval through
\begin{align}
\sum_{i \in \tilde{\cal F}_{\leq q-2}} \gamma_i \cdot I({\cal P}^{B*}_i, b_\mathrm{exit}^-) & ~~=~~ \sum_{i \in \tilde{\cal F}_{\leq q-2}} \gamma_i \cdot \left( I({\cal P}^{B*}_i, b_\mathrm{entry}^+) - (b_\mathrm{exit} - b_\mathrm{entry}) \right)  \nonumber \\
& ~~=~~ \sum_{i \in \tilde{\cal F}_{\leq q-2}} \gamma_i \cdot  I({\cal P}^{B*}_i, b_\mathrm{entry}^+) - \frac{ \tau_\mycyc }{ (n/\eps)^{3q-4} } \cdot \sum_{i \in \tilde{\cal F}_{\leq q-2}} \gamma_i \ , \label{eqn:recursive_before_after}
\end{align}
where the second equality holds since $b_\mathrm{exit} - b_\mathrm{entry} = \frac{ \tau_\mycyc }{ |B_q^-| } = \frac{ \tau_\mycyc }{ (n/\eps)^{3q-4} }$. Given this observation, rather than accurately keeping track of $\{ I({\cal P}^{B*}_i, b_\mathrm{entry}^+) \}_{i \in \tilde{\cal F}_{\leq q-2}}$, our proxy statistic will form a lower bound on the total space requirement of these commodities at any point in $[b_\mathrm{entry}, b_\mathrm{exit})$. This statistic, termed ``$\tilde{\cal F}_{\leq q-2}$-space bound'', is given by $\sum_{i \in \tilde{\cal F}_{\leq q-2}} \gamma_i \cdot  I({\cal P}^{B*}_i, b_\mathrm{exit}^-)$, which is exactly the left-hand side of equation~\eqref{eqn:recursive_before_after}. Interestingly, our dynamic program will falsely assume that the space requirement of these commodities remains unchanged at this statistic across $[b_\mathrm{entry}, b_\mathrm{exit})$. While this false assumption may lead to capacity violations, we will prove that the maximum possible violation is negligible.

\subsection{The dynamic program: Guessing and state space} \label{subsec:dp_guess_State}

\paragraph{Guessing.} Having laid the theoretical foundations of our algorithmic approach, we move on to describe its concrete implementation. For simplicity of presentation, we initially  guess two ingredients that appeared in earlier sections, while being unknown from an algorithmic standpoint up until now:
\begin{itemize}
    \item {\em Cycle length $\tau_\mycyc$}: We first obtain an over-estimate $\tilde{\tau}_\mycyc$ for the cycle length $\tau_\mycyc$ within a factor of $1 + \eps$, meaning that $\tilde{\tau}_\mycyc \in [\tau_\mycyc, (1+\eps) \cdot \tau_\mycyc]$. Given Theorem~\ref{thm:good_cyclic}(2), we know that $\tau_{\mycyc} \in [\frac{ K_{\max} }{ 2\eps n {\cal M} }, \frac{ 2 n {\cal M} }{ \eps^2 H_{\min} }]$, meaning that the number of values to be tested as possible estimates is $O( \log_{ 1 + \eps } ( \frac{ 4n^2 {\cal M}^2 }{ \eps K_{\max}  H_{\min} } ) ) = O( (\frac{ |{\cal I}| }{ \eps } )^{ O(1) })$. For convenience, we make direct use of $\tau_\mycyc$; the negligible effects of this simplification will be discussed as part of our analysis.

    \item {\em Frequency classes $\tilde{\cal F}_1, \ldots, \tilde{\cal F}_Q$}: Second, for each commodity $i \in [n]$, we guess the frequency class to which it belongs, leading to $Q^n$ options to be enumerated over. As mentioned in Section~\ref{subsec:structure_thm}, we have $Q = \lceil \log_{(n/\eps)^{3}} {\cal U}  \rceil$ and ${\cal U} = \frac{ 4 n^2 {\cal M}^2 }{ \eps^2 K_{\min} H_{\min} }$, implying that $Q^n = O( (\frac{ |{\cal I}| }{ \eps } )^n)$.
\end{itemize}

\begin{figure}[htbp!]
    \centering
    \resizebox{0.8\textwidth}{!}{
    \begin{tikzpicture}[scale=1.5, >=latex]
        \def\bentry{0}
        \def\bexit{7.5}
        \def\ymax{5}
        \def\floorh{1.5}

        \fill[gray!20] (\bentry,0) rectangle (\bexit,\floorh);
        \draw[thick, gray!80] (\bentry,\floorh) -- (\bexit,\floorh);

        \node at (3.75, 0.75) [align=center, font=\small] {$\tilde{\mathcal{F}}_{\leq q-2}$-space bound $\mathcal{L}_{\leq q-2}$ \\ (classes $1, \dots, q-2$)};

        \draw[blue, thick]
            (0.0, 2.2) -- (0.5, 1.7) -- (0.5, 2.5) 
            -- (1.0, 2.0) -- (1.0, 2.3)           
            -- (1.5, 1.8) -- (1.5, 2.8)           
            -- (2.0, 2.3) -- (2.0, 2.6)           
            -- (2.5, 2.1) -- (2.5, 2.4)           
            -- (3.0, 1.9) -- (3.0, 2.7)           
            -- (3.5, 2.2) -- (3.5, 2.5)           
            -- (4.0, 2.0) -- (4.0, 2.3)           
            -- (4.5, 1.8) -- (4.5, 2.1)           
            -- (5.0, 1.6) -- (5.0, 2.5)           
            -- (5.5, 2.0) -- (5.5, 2.3)           
            -- (6.0, 1.8) -- (6.0, 2.2)           
            -- (6.5, 1.7) -- (6.5, 2.7)           
            -- (7.0, 2.2) -- (7.0, 2.5)           
            -- (7.5, 2.0) -- (7.5, 2.3);          

        \foreach \x/\y in {
            0.0/2.2, 0.5/2.5, 1.0/2.3, 1.5/2.8,
            2.0/2.6, 2.5/2.4, 3.0/2.7, 3.5/2.5,
            4.0/2.3, 4.5/2.1, 5.0/2.5, 5.5/2.3,
            6.0/2.2, 6.5/2.7, 7.0/2.5, 7.5/2.3
        } {
            \filldraw[blue] (\x,\y) circle (1.5pt);
        }

        \node[blue] at (2.5, 3.5) [font=\small, align=center] {$\tilde{\mathcal{F}}_{q-1}$-inventory levels $\mathcal{I}_{q-1}$ \\ (class $q-1$)};

        \draw[->, thick] (0,0) -- (8.0,0) node[right, font=\small] {$t$};
        \draw[->, thick] (0,0) -- (0,5.5) node[above, font=\small] {space};

        \draw[thick, dashed] (0,5) -- (7.5,5);
        \node at (7.5, 5.0) [anchor=south east, font=\small] {$\mathcal{V}$ (max capacity)};

        \draw (0,0.1) -- (0,-0.1) node[below, font=\small] {$b_{\text{entry}}$};
        \draw (7.5,0.1) -- (7.5,-0.1) node[below, font=\small] {$b_{\text{exit}}$};

        \foreach \i in {1, ..., 14} {
            \pgfmathsetmacro{\pos}{\i * 0.5}
            \draw (\pos, 0.05) -- (\pos, -0.05);
        }

    \end{tikzpicture}
    }
    \caption{Visual representation of our state description { $(q, {\cal I}_{q-1}, \mylb_{ \leq q-2 })$}. The gray region corresponds to the space reserved by lower-frequency commodities. The blue sawtooth curve captures the aggregate inventory levels across all { $\tilde{\cal F}_{q-1}$}-commodities, which are replenished at { $B_{q-1}^+$}-points.}
    \label{fig:dp_state_layered}
\end{figure}
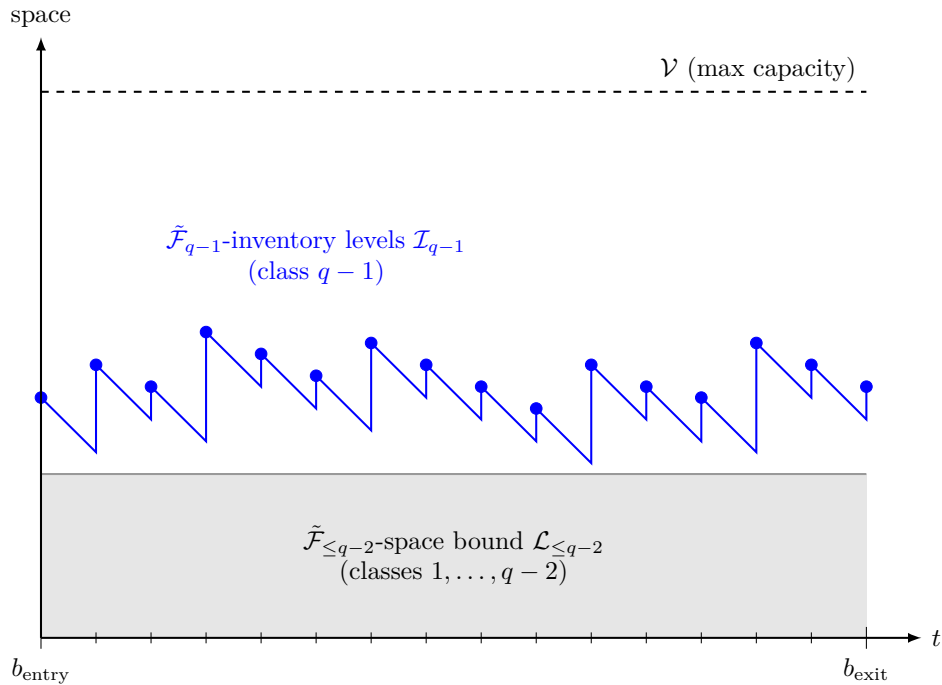

\paragraph{State space.} Each state $(q, {\cal I}_{q-1}, \mylb_{ \leq q-2 })$ of our dynamic program consists of the next three parameters, which are schematically illustrated in Figure~\ref{fig:dp_state_layered}:
\begin{enumerate}
    \item {\em Class index $q$}: This index corresponds to the frequency class $\tilde{\cal F}_q$ for which we are currently placing orders, across an interval stretching between two successive $B_q^-$-points. We emphasize that this interval is not explicitly specified within our state description. Rather, in conjunction with ${\cal I}_{q-1}$ and $\mylb_{ \leq q-2 }$, we will employ the exact same policy for any interval between two successive $B_q^-$-points; for clarity, we symbolically denote such an interval by $[b_\mathrm{entry}, b_\mathrm{exit}]$. It is important to point out that only non-empty frequency classes are ever considered by our dynamic program, meaning that $q$ will be taking $O( n )$ possible values out of $1, \ldots, Q$, even though $Q$ itself could be much larger.

    \item {\em $\tilde{\cal F}_{q-1}$-inventory levels ${\cal I}_{q-1}$}: Let $x_1, \ldots, x_R$ be the sequence of $(\frac{n}{\eps})^2$ points in $B_{q-1}^+$ where $B$-aligned policies are allowed to place $\tilde{\cal F}_{q-1}$-orders along $[b_\mathrm{entry}, b_\mathrm{exit}]$. That is,
    \[ x_1 ~~=~~ b_\mathrm{entry}, \qquad x_2 ~~=~~ b_\mathrm{entry} + \frac{ b_\mathrm{exit} - b_\mathrm{entry} }{ ( n/\eps )^2 }, \qquad \ldots, \qquad x_R ~~=~~ b_\mathrm{exit} \ . \]
    Then, the $(|\tilde{\cal F}_{q-1}| \cdot R)$-dimensional  vector ${\cal I}_{q-1}$ will represent the already-decided inventory level of each commodity $i \in \tilde{\cal F}_{q-1}$ at each of the points $x_1, \ldots, x_R$. As explained in Section~\ref{subsec:approx_stat}, this vector can take $O ( 2^{ O( n^3 / \eps^2 ) } )$ different values.

    \item {\em $\tilde{\cal F}_{\leq q-2}$-space bound $\mylb_{ \leq q-2 }$}: This parameter serves as a lower bound on the already-decided total space requirement of $\tilde{\cal F}_{\leq q-2}$-commodities at any point in  $[b_\mathrm{entry}, b_\mathrm{exit})$. As explained in Section~\ref{subsec:recursive_details}, we will restrict $\mylb_{ \leq q-2 }$ to integer multiples of $\frac{ \eps }{ n } \cdot {\cal V}$ within $[0, (1 + \eps) \cdot {\cal V}]$, meaning that this parameter only takes $O( \frac{ n }{ \eps } )$ values.
\end{enumerate}
In light of this description, the overall number of states  is $O(n \cdot 2^{ O( n^3 / \eps^2 ) } \cdot \frac{ n }{ \eps } ) = O( 2^{ O( n^3 / \eps^2 ) } )$.

\subsection{The dynamic program: Recursive equations} \label{subsec:recursive_details}

In what follows, we describe how our dynamic program operates at each state $(q, {\cal I}_{q-1}, \mylb_{ \leq q-2 })$. Broadly speaking, this procedure will compute a $B$-aligned policy for all $\tilde{\cal F}_{ \geq q }$-commodities over an interval $[b_\mathrm{entry}, b_\mathrm{exit})$ that stretches between two successive $B_q^-$-points, assuming that our actions regarding $\tilde{\cal F}_{\leq q-1}$-commodities resulted in $\tilde{\cal F}_{q-1}$-inventory levels ${\cal I}_{q-1}$ and $\tilde{\cal F}_{\leq q-2}$-space bound $\mylb_{ \leq q-2 }$ for this interval. We make use of $C(q, {\cal I}_{q-1}, \mylb_{ \leq q-2 })$ to denote the total cost of this policy across $[b_\mathrm{entry}, b_\mathrm{exit})$.

\paragraph{Step 1: Action space for {\boldmath ${\tilde{\cal F}_q}$}.} Let us begin by observing that, given property~\ref{item:B_aligned_2} of $B$-aligned policies, for every commodity $i \in \tilde{\cal F}_q$, such policies are allowed to place $i$-orders only at $B_q^+$-points. Since successive $B_q^+$-points are positioned within a distance $\frac{ \tau_\mycyc }{ | B_q^+| }$ of each other, and since $[b_\mathrm{entry}, b_\mathrm{exit}]$ is of length $\frac{ \tau_\mycyc }{ | B_q^-| }$, it follows that the number of $B_q^+$-points in this interval is $\frac{ | B_q^+ | }{ |B_q^-| } = ( \frac{ n }{ \eps } )^{ 5 }$. For future reference, we use $y_1, \ldots, y_S$ to denote these points in left-to-right order. Consequently, jointly over all commodities in $\tilde{\cal F}_q$, there are only $2^{ |\tilde{\cal F}_q| \cdot S } = 2^{ O( n^{6} / \eps^{5} ) }$ possible $B$-aligned policies to be considered. For each such policy ${\cal P}^q$, we go through steps~2 and~3 below, where ${\cal P}^q$ will either be recursively extended to a $B$-aligned policy ${\cal P}^{\geq q}$ for all $\tilde{\cal F}_{ \geq q }$-commodities or tagged as being ``unextendable''. In the event that all policies tested are unextendable, our algorithm will report this fact by returning $C(q, {\cal I}_{q-1}, \mylb_{ \leq q-2 }) = \bot$. In the opposite scenario, we return the extendable policy ${\cal P}^{\geq q}$ whose cost $C( {\cal P}^{\geq q}, [b_\mathrm{entry}, b_\mathrm{exit}))$ is minimal, setting $C(q, {\cal I}_{q-1}, \mylb_{ \leq q-2 })$ to this value.

\paragraph{Step 2: Verifying acceptability.} Due to limiting our attention to $B$-aligned policies, once the vector ${\cal I}_{q-1}$ of $\tilde{\cal F}_{q-1}$-inventory levels is fixed, it determines a unique replenishment policy ${\cal P}^{q-1}$ for the $\tilde{\cal F}_{q-1}$-commodities across $[b_\mathrm{entry}, b_\mathrm{exit})$. In turn, we say that the policy ${\cal P}^q$ is acceptable when, for every point $t \in [b_\mathrm{entry}, b_\mathrm{exit})$,
\begin{equation} \label{eqn:def_acceptable}
\underbrace{ \sum_{i \in \tilde{\cal F}_q} \gamma_i \cdot I( {\cal P}^q_i, t ) }_{ \text{exact space due to $\tilde{\cal F}_q$} } + \underbrace{ \sum_{i \in \tilde{\cal F}_{q-1}} \gamma_i \cdot I( {\cal P}^{ q-1}_i, t ) }_{ \text{exact space due to $\tilde{\cal F}_{q-1}$} } + \underbrace{ \vphantom{\sum_{i \in \tilde{\cal F}_{q-1}}} \mylb_{ \leq q-2 } }_{ \MyAbove{ \text{lower bound on} }{ \text{space due to $\tilde{\cal F}_{\leq q-2}$} } } ~~\leq~~ (1 +\eps) \cdot {\cal V} \ .
\end{equation}
It is important to point out that this definition is weaker than $(1 +\eps)$-feasibility, since $\mylb_{ \leq q-2 }$ is only a lower bound on the space requirement of $\tilde{\cal F}_{\leq q-2}$-commodities. Another important observation is that, even though condition~\eqref{eqn:def_acceptable} is written for all points $t \in [b_\mathrm{entry}, b_\mathrm{exit})$, we can verify whether ${\cal P}^q$ is acceptable in polynomial time. Specifically, it suffices to examine whether this condition holds for the $B_q^+$-points in this interval, $y_1, \ldots, y_S$, since the inventory level of any commodity in $\tilde{\cal F}_q \cup \tilde{\cal F}_{q-1}$ is decreasing between any pair of successive points, $y_s$ and $y_{s+1}$.  We proceed to step~3 only when ${\cal P}^q$ is acceptable; otherwise, this policy is eliminated from being further considered, and we tag it as unextendable. In what follows, step~3A considers the case where $\tilde{\cal F}_{q+1} \neq \emptyset$, while step~3B discusses the one where $\tilde{\cal F}_{q+1} = \emptyset$.

\paragraph{Step 3A: Recursive policy for {\boldmath ${\tilde{\cal F}_{\geq q+1}}$}, when {\boldmath ${\tilde{\cal F}_{q+1} \neq \emptyset}$}.} Noting that $|B_{q+1}^-| = (\frac{n}{\eps})^{3(q+1) - 4} $ and $| B_q^-| = (\frac{n}{\eps})^{3q-4}$, it follows that $[b_\mathrm{entry}, b_\mathrm{exit})$ is subdivided by $B_{q+1}^-$-points into $( \frac{n}{\eps})^3$ equal-length intervals, say $I_1, \ldots, I_M$ in left-to-right order.  Let us focus on a single  interval $I_m = [b_\mathrm{entry}^m, b_\mathrm{exit}^m)$, obviously belonging to the next level of our recursion, $q+1$, which indeed exists, since $\tilde{\cal F}_{q+1} \neq \emptyset$ by the hypothesis of step~3A. The important observation is that, since we have already fixed the policy ${\cal P}^q$, it uniquely determines the vector ${\cal I}_q^m$ of $\tilde{\cal F}_q$-inventory levels  associated with this interval. In addition, since the policy ${\cal P}^{q-1}$ is fixed as well (see step~2), we obtain a lower bound $\mylb_{ \leq q-1 }^m$ on the total space requirement of $\tilde{\cal F}_{\leq q-1}$-commodities at any point in  $I_m = [b_\mathrm{entry}^m, b_\mathrm{exit}^m)$ by setting
\begin{equation} \label{eqn:recur_lb_case1}
\mylb_{ \leq q-1 }^m ~~=~~ \mylb_{ \leq q-2 } + \left\lfloor \sum_{i \in \tilde{\cal F}_{q-1} } \gamma_i \cdot I( {\cal P}^{q-1}_i, b_\mathrm{exit}^{m-}) \right\rfloor_{ \eps {\cal V} / n} \ .
\end{equation}
Here, $\lfloor \cdot \rfloor_{ \eps {\cal V} / n}$ is an operator that rounds its argument down to the nearest integer multiple of $\frac{ \eps }{ n } \cdot {\cal V}$.

Next, for every interval $I_m$, we examine the actions taken by our dynamic program at state $(q+1, {\cal I}_q^m, \mylb_{ \leq q-1 }^m)$. When $C(q+1, {\cal I}_q^m, \mylb_{ \leq q-1 }^m) = \bot$ for some $m \in [M]$, we report that ${\cal P}^q$ is unextendable. In the opposite case, for every $m \in [M]$, we obtain a $B$-aligned policy ${\cal P}^{\geq q+1,m}$ for all $\tilde{\cal F}_{ \geq q+1 }$-commodities over $I_m$, with cost $C( {\cal P}^{\geq q+1, m}, I_m ) = C(q+1, {\cal I}_q^m, \mylb_{ \leq q-1 }^m)$. Finally, we create a $B$-aligned policy ${\cal P}^{ \geq q}$ for all $\tilde{\cal F}_{ \geq q }$-commodities by gluing ${\cal P}^q$ together with the concatenation of ${\cal P}^{\geq q+1,1}, \ldots, {\cal P}^{\geq q+1,M}$ in this order.

\paragraph{Step 3B: Recursive policy for {\boldmath ${\tilde{\cal F}_{\geq q+1}}$}, when {\boldmath ${\tilde{\cal F}_{q+1} = \emptyset}$}.} Again, let us focus on a single interval $I_m = [b_\mathrm{entry}^m, b_\mathrm{exit}^m)$. In contrast to step~3A, we cannot employ recursive calls to level $q+1$, since our dynamic program does not have states of the form $(q+1, \cdot, \cdot)$ when $\tilde{\cal F}_{q+1} = \emptyset$, which is exactly the hypothesis of step~3B. Toward circumventing this issue, let $q_\mynext$ be the minimum index of a non-empty frequency class out of $\tilde{\cal F}_{q+2}, \ldots, \tilde{\cal F}_{Q}$. Rather than descending to the non-existing level $q+1$, we will proceed directly to level $q_\mynext$. Technically speaking, we fill $I_m$ by gluing identical copies of the policy ${\cal P}^{ \geq q_\mynext, m }$, returned by our dynamic program at the single state $(q_\mynext, {\cal I}_{q_\mynext - 1}^m, \mylb_{ \leq q_\mynext-2 }^m)$, where ${\cal I}_{q_\mynext - 1}^m$ and $\mylb_{ \leq q_\mynext-2 }^m$ are determined as follows:
\begin{itemize}
    \item Since $\tilde{\cal F}_{q+1} = \emptyset$ by the case hypothesis, we know that $\tilde{\cal F}_{q_\mynext-1} = \emptyset$ as well. Consequently, our vector  ${\cal I}_{q_\mynext - 1}^m$ of $\tilde{\cal F}_{q_\mynext-1}$-inventory levels does not  exist (equivalently, it is the zero vector), which will be indicated by ${\cal I}_{q_\mynext - 1}^m = \vec{0}$.

    \item Along the lines of step~3A, our lower bound $\mylb_{ \leq q_\mynext-2 }^m$ on the total space requirement of $\tilde{\cal F}_{\leq q_\mynext-2}$-commodities at any point will be
    \begin{equation} \label{eqn:recur_lb_case2}
    \mylb_{ \leq q_\mynext-2 }^m ~~=~~ \mylb_{ \leq q-2 } + \left\lfloor \sum_{i \in \tilde{\cal F}_q } \gamma_i \cdot I( {\cal P}^{q}_i, b_\mathrm{exit}^{m-}) + \sum_{i \in \tilde{\cal F}_{q-1} } \gamma_i \cdot I( {\cal P}^{q-1}_i, b_\mathrm{exit}^{m-}) \right\rfloor_{ \eps {\cal V} / n} \ .
    \end{equation}
\end{itemize}
Of course, when $C(q_\mynext, {\cal I}_{q_\mynext - 1}^m, \mylb_{ \leq q_\mynext-2 }^m) = \bot$, we report that ${\cal P}^q$ is unextendable. It is important to notice that, in order to fill $I_m$, the number of ${\cal P}^{ \geq q_\mynext, m }$-copies required is $\frac{ | B_{q_\mynext}^-| }{ | B_{q-1}^-| } = ( \frac{ n }{ \eps } )^{ 3(q_\mynext - q+1) }$. This quantity can be as large as $( \frac{ n }{ \eps } )^{\Omega(Q)}$, which is precisely the type of running time dependency we wish to avoid. However, since we  make use of identical copies, only one needs to be computed and represented. Finally, as in step~3A, we create a $B$-aligned policy ${\cal P}^{ \geq q}$ for all $\tilde{F}_{ \geq q }$-commodities by gluing ${\cal P}^q$ together with the concatenation of the above-mentioned policies for $I_1, \ldots, I_M$.

\paragraph{Top level.} To describe the overall policy we return for the entire cycle $[0, \tau_\mycyc)$, it is convenient to assume that $\tilde{\cal F}_1 \neq \emptyset$. This assumption is without loss of generality, since we can add a dummy commodity, indexed $0$, parameterized by $H_0 = K_0 = \gamma_0 = 0$, which is ordered by the policy $\tilde{\cal P}$ so that $0 \in \tilde{\cal F}_1$. As such, our overall policy is the one computed at state $(1, \vec{0}, 0)$.

\subsection{Analysis: Computability and performance guarantees} \label{subsec:DP_analysis}

In what follows, we begin by addressing the most basic question: Regardless of capacity-feasibility and cost, will our dynamic program even compute a replenishment policy for the entire cycle? As explained below, this hidden question is a-priori unclear. Subsequently, letting ${\cal P}^{ \mydp }$ be the resulting policy, we will show that it indeed meets the performance guarantees stated in the opening paragraph of Section~\ref{subsec:recusive_main_result}. In other words, we will prove that ${\cal P}^{ \mydp }$ is $(1 + O(\eps))$-feasible and that its cost is $C( {\cal P}^{ \mydp }, [0,\tau_{\mycyc}) ) = (1+O(\eps)) \cdot C( \hat{\cal P}, [0,\tau_{\mycyc}) )$. Finally, we will discuss the running time of our particular implementation.

\paragraph{Extendability and cost.} Clearly,
our dynamic program computes a replenishment policy for the entire cycle if and only if  $C(1, \vec{0}, 0) \neq \bot$. The crucial observation is that, by going through each and every algorithmic step of Section~\ref{subsec:recursive_details}, it is not difficult to verify that the optimal $B$-aligned $(1+\eps)$-feasible policy ${\cal P}^{B*}$ for the segment $[0,\tau_{\mycyc})$ is a feasible solution to our recursion. In other words, at level $q=1$, one of the policies examined in step~1 as part of the action space for $\tilde{\cal F}_1$ is duplicating the orders placed by $\{ {\cal P}^{B*}_i \}_{ i \in \tilde{\cal F}_1 }$, which obviously passes our acceptability check in step~2, since this condition is weaker than $(1+\eps)$-feasibility.  Once we descend to the second recursion level (say $q=2$) with this policy, one of the actions examined is duplicating $\{ {\cal P}^{B*}_i \}_{ i \in \tilde{\cal F}_2 }$, so on and so forth. Consequently, we are guaranteed to have $C(1, \vec{0}, 0) \neq \bot$, implying that our dynamic program indeed computes a replenishment policy ${\cal P}^{ \mydp }$ for the entire cycle. Moreover, in terms of cost, these observations supposedly imply that $C( {\cal P}^{ \mydp }, [0,\tau_{\mycyc}) ) \leq C( {\cal P}^{B*}, [0,\tau_{\mycyc}) )$. However, due to making direct use of $\tau_{\mycyc}$ in place of its estimate $\tilde{\tau}_\mycyc \in [\tau_\mycyc, (1+\eps) \cdot \tau_\mycyc]$, we actually have
\[ C( {\cal P}^{ \mydp }, [0,\tau_{\mycyc}) ) ~~\leq~~ (1 + \eps) \cdot C( {\cal P}^{B*}, [0,\tau_{\mycyc}) ) ~~\leq~~ (1 + \eps) \cdot
 C( \hat{\cal P}, [0,\tau_{\mycyc}) ) \ , \]
where the second inequality holds since, by Theorem~\ref{thm:exist_aligned}, we know that $\hat{\cal P}$ is a $B$-aligned $(1+\eps)$-feasible policy for the segment $[0,\tau_{\mycyc})$, and since ${\cal P}^{B*}$ is a minimum-cost such policy.

\paragraph{Space requirement.} The next question we address is regarding the peak space requirement of ${\cal P}^{ \mydp }$, noting that our dynamic program does not explicitly enforce capacity-feasibility. The following claim shows that ${\cal P}^{ \mydp }$ is $(1 + 2\eps)$-feasible.

\begin{lemma}
$V_{\max}( {\cal P}^{ \mydp } ) \leq (1 + 2\eps) \cdot {\cal V}$.
\end{lemma}
\begin{proof}
To bound the space requirement of ${\cal P}^{ \mydp }$ at any point in $[0,\tau_{\mycyc})$, suppose that throughout our dynamic program, we would have augmented each state $(q, {\cal I}_{q-1}, \mylb_{ \leq q-2 })$ with a fourth parameter, $\myub_{ \leq q-2 }$. Operating in the opposite direction of the lower bound $\mylb_{ \leq q-2 }$, this parameter would have served as an upper bound on the already-decided total space requirement of $\tilde{\cal F}_{\leq q-2}$-commodities at any point in  $[b_\mathrm{entry}, b_\mathrm{exit})$. To recursively update this parameter, one can mimic equations~\eqref{eqn:recur_lb_case1} and~\eqref{eqn:recur_lb_case2}, such that:
\[ \begin{cases}
    \myub_{ \leq q-1 }^m ~~=~~ \myub_{ \leq q-2 } +  \sum_{i \in \tilde{\cal F}_{q-1} } \gamma_i \cdot I( {\cal P}^{q-1}_i, b_\mathrm{entry}^{m+}), & \text{when } \tilde{\cal F}_{q+1} \neq \emptyset \\
    \myub_{ \leq q_\mynext-2 }^m ~~=~~ \myub_{ \leq q-2 } +  \sum_{i \in \tilde{\cal F}_q } \gamma_i \cdot I( {\cal P}^{q}_i, b_\mathrm{entry}^{m+}) + \sum_{i \in \tilde{\cal F}_{q-1} } \gamma_i \cdot I( {\cal P}^{q-1}_i, b_\mathrm{entry}^{m+}), & \text{when } \tilde{\cal F}_{q+1} = \emptyset
\end{cases}\]
Now, by combining this definition with equation~\eqref{eqn:recursive_before_after}, we conclude that $\myub_{ \leq q-2 }^m - \mylb_{ \leq q-2 }^m \leq \frac{ \tau_\mycyc }{ (n/\eps)^{3q-4} } \cdot \sum_{i \in \tilde{\cal F}_{\leq q-2}} \gamma_i + \frac{ \eps {\cal V} }{ n }$, for every class index $q \geq 3$ with $\tilde{\cal F}_q \neq \emptyset$. As such, the maximum-possible space violation of ${\cal P}^{ \mydp }$ across the cycle $[0,\tau_{\mycyc})$ is at most
\begin{align}
\sum_{q \geq 3: \tilde{\cal F}_q \neq \emptyset } \left( \myub_{ \leq q-2 }^m - \mylb_{ \leq q-2 }^m  \right) & ~~\leq~~ \sum_{q \geq 3: \tilde{\cal F}_q \neq \emptyset } \frac{ \tau_\mycyc }{ (n/\eps)^{3q-4} } \cdot \sum_{i \in \tilde{\cal F}_{\leq q-2}} \gamma_i + \eps {\cal V} \nonumber \\
& ~~\leq~~ \left( \frac{ \eps }{ n } \right)^2 \cdot {\cal V} \cdot \sum_{q \geq 3: \tilde{\cal F}_q \neq \emptyset } | \tilde{\cal F}_{\leq q-2} |  + \eps {\cal V} \label{eqn:rec_space_bound} \\
& ~~\leq~~ \frac{ \eps^2 }{ n } \cdot {\cal V} + \eps {\cal V} \nonumber \\
& ~~\leq~~ 2\eps {\cal V} \ . \nonumber
\end{align}
Here, inequality~\eqref{eqn:rec_space_bound} is obtained by arguing that $\gamma_i \cdot \frac{ \tau_\mycyc }{ (n/\eps)^{3(q-2)} } \leq {\cal V}$ for every commodity $i \in \tilde{\cal F}_{\leq q-2}$. To this end, by definition of $\tilde{\cal F}_1, \ldots, \tilde{\cal F}_{q-2}$, we know that with respect to the policy $\tilde{\cal P}_i$, this commodity has $N( \tilde{\cal P}_i, [0,\tau_{\mycyc})) \leq (\frac{n}{\eps})^{3(q-2)}$ orders across the entire cycle. Therefore, at least one of these orders consists of at least $\frac{ \tau_\mycyc }{ (n/\eps)^{3(q-2)} }$ units, which require $\gamma_i \cdot \frac{ \tau_\mycyc }{ (n/\eps)^{3(q-2)} }$ amount of space by themselves. As shown in Theorem~\ref{thm:good_cyclic}, the policy $\tilde{\cal P}$ is capacity-feasible, implying in particular that $\gamma_i \cdot \frac{ \tau_\mycyc }{ (n/\eps)^{3(q-2)} } \leq {\cal V}$.
\end{proof}

\paragraph{Running time.} As far as efficiency is concerned, as explained in Section~\ref{subsec:dp_guess_State}, our preliminary guessing step involves enumerating over $O( (\frac{ |{\cal I}| }{ \eps } )^n)$ possible configurations of the cycle length estimate $\tilde{\tau}_{ \mycyc }$ and the frequency classes $\tilde{\cal F}_1, \ldots, \tilde{\cal F}_Q$. Given these guesses, our dynamic program consists of $O( 2^{ O( n^3 / \eps^2 ) } )$ states. Following the discussion in Section~\ref{subsec:recursive_details}, at each of these states, there are only $2^{ O( n^{6} / \eps^{5} ) }$ possible $B$-aligned policies to be examined. Any such policy can be tested for acceptability in polynomial time, and subsequently extended via $( \frac{n}{\eps})^3$ recursive calls. Putting these pieces together, a straightforward implementation requires $O( | {\cal I} |^{O(n)} \cdot 2^{ O( n^{6} / \eps^{5} ) } )$ time overall.

\section{Concluding Remarks} \label{sec:conclusions}

In addition to resolving open computational questions about the economic warehouse lot scheduling problem, bypassing inherent algorithmic challenges and polynomial-space policy representability, our work has  played an  important role in breaking the long-standing $2$-approximation barrier for general problem instances. The next few paragraphs are intended to succinctly discuss this outcome, as well as to single out prospective directions for future research.

\paragraph{Sub-{\boldmath $2$}-approximation for general instances.} In subsequent work~\citep{Segev26EWLSPsub2}, we developed new analytical ideas and algorithmic advances, culminating in a proof-of-concept improvement on the long-standing performance guarantees of \cite{Anily91} and \cite{GallegoQS96} for general problem instances, consisting of arbitrarily-many commodities. Specifically, for any $\eps > 0$, we devised an $O( | {\cal I} |^{\tilde{O}( 1/\eps^5)} \cdot 2^{ \tilde{O}( 1 / \eps^{35} ) } )$-time randomized construction of a capacity-feasible policy whose expected long-run average cost is within factor $2-\frac{17}{5000} + \eps$ of optimal. Once again, it is important to point out that these developments are by no means attempting to
arrive at either the best-possible approximation ratio or the most efficient implementation. Interested readers are referred to the resulting paper~\citep{Segev26EWLSPsub2} for additional details on this long-awaited improvement as well as on why our approximation scheme for constantly-many commodities seems absolutely essential.

\paragraph{The strategic version.} As pointed out by \citet[p.~51]{GallegoSS92}, and as one can discover by reviewing adjacent literature, the vast majority of academic research around economic warehouse lot scheduling has been dedicated to its  ``tactical'' version, which is precisely the one studied in this paper. Yet, considerable attention has concurrently been given to the so-called ``strategic'' version (see, e.g., \cite{HodgsonL82, ParkY85, Hall88, RosenblattR90}), where  rather than having a warehouse space constraint, this term appears as part of the objective function; namely, we wish to minimize $C({\cal P}) + V_{\max}( {\cal P} )$. It is not difficult to verify that, by guessing the $V_{\max}( {\cal P}^* )$-term of an optimal policy ${\cal P}^*$ and writing an appropriate space constraint, we can show that a tactical $\alpha$-approximation can be converted to a strategic $(\sqrt{\alpha} + \eps)$-approximation. Consequently, our  work provides a polynomial-time approximation scheme for constantly-many commodities in this context. As an avenue for future research, given the extra leeway of having $V_{\max}( {\cal P} )$ within the objective function rather than within a hard space constraint, it will be interesting to examine whether the strategic version admits a qualitatively simpler approximation scheme. Despite our best efforts, this question remains wide open at present time.

\paragraph{Efficient implementation?} As formally stated in Theorem~\ref{thm:PTAS_exponential}, our main algorithmic finding is captured by an $O( | {\cal I} |^{O(n)} \cdot 2^{ O( n^{6} / \eps^{5} ) } )$-time dynamic programming approach for approximating the economic warehouse lot scheduling problem within factor $1 + \eps$ of optimal. Along these lines, much of the running time exponent can be attributed to avoiding a highly-optimized implementation of our approximation scheme, mainly for clarity of exposition and readability. Even though we have not attempted to eliminate such dependencies, there may be an alternative way to synthesize our main ideas, ending up with a  more efficient implementation. Of course, the potential inaccessibility of efforts in this spirit is still unknown.

\phantomsection
\addcontentsline{toc}{section}{References}
\bibliographystyle{plainnat}
\bibliography{BIB-Lot-Sizing}

@misc{Segev26EWLSPsub2,
    author = {Danny Segev},
    title = {Economic Warehouse Lot Scheduling: Breaking the $2$-Approximation Barrier},
    year = {2026},
    note = {Working paper, submitted to publication}
}

@book{NahmiasL15,
  title={Production and Operations Analysis},
  author={Nahmias, Steven and Olsen, Tava Lennon},
  year={2015},
  publisher={Waveland Press}
}

@book{HaxC84,
  title={Production and Inventory Management},
  author={Arnoldo C. Hax and Dan Candea},
  publisher={Prentice-Hall},
  year={1984}
}

@book{Vazirani01,
  title={Approximation Algorithms},
  author={Vazirani, Vijay V.},
  year={2001},
  publisher={Springer}
}

@book{JohnsonM74,
  title={Operations Research in Production Planning, Scheduling, and Inventory Control},
  author={Johnson, Lynwood A. and Montgomery, Douglas C.},
  publisher={Wiley},
  year={1974}
}

@book{HadleyW1963,
  title={Analysis of Inventory Systems},
  author={George Hadley and Thomson M. Whitin},
  year={1963},
  publisher={Prentice-Hall}
}

@article{GallegoQS96,
  author       = {Guillermo Gallego and
                  Maurice Queyranne and
                  David Simchi{-}Levi},
  title        = {Single Resource Multi-Item Inventory Systems},
  journal      = {Operations Research},
  volume       = {44},
  number       = {4},
  pages        = {580--595},
  year         = {1996}
}

@book{SimchiLeviCB14,
  title={The Logic of Logistics: Theory, Algorithms, and Applications for Logistics Management},
  author={David Simchi-Levi and Xin Chen and Julien Bramel},
  publisher={Springer Science \& Business Media},
  year={2014},
  edition={third}  
}

@article{Anily91,
  author       = {Shoshana Anily},
  title        = {Multi-Item Replenishment and Storage Problem {(MIRSP):} {H}euristics and Bounds},
  journal      = {Operations Research},
  volume       = {39},
  number       = {2},
  pages        = {233--243},
  year         = {1991}
}

@article{HodgsonL82,
  title={Production lot sizing with material-handling cost considerations},
  author={Hodgson, Thom J. and Lowe, Timothy J.},
  journal={IIE Transactions},
  volume={14},
  number={1},
  pages={44--51},
  year={1982}
}

@article{RosenblattR90,
  title={On the single resource capacity problem for multi-item inventory systems},
  author={Rosenblatt, Meir J. and Rothblum, Uriel G.},
  journal={Operations Research},
  volume={38},
  number={4},
  pages={686--693},
  year={1990}
}

@article{Hall88,
  title={A multi-item {EOQ} model with inventory cycle balancing},
  author={Hall, Nicholas G.},
  journal={Naval Research Logistics},
  volume={35},
  number={3},
  pages={319--325},
  year={1988}
}

@article{ParkY85,
  title={Optimal scheduling of periodic activities},
  author={Park, Kyung S. and Yun, Doek K.},
  journal={Operations Research},
  volume={33},
  number={3},
  pages={690--695},
  year={1985}
}

@article{Holt58,
  title={Decision rules for allocating inventory to lots and cost foundations for making aggregate inventory decisions},
  author={Holt, Charles C.},
  journal={Journal of Industrial Engineering},
  volume={9},
  pages={14--22},
  year={1958}
}

@article{Homer1966,
  title={Space-limited aggregate inventories with phased deliveries},
  author={Homer, Eugene D.},
  journal={Journal of Industrial Engineering},
  volume={17},
  number={6},
  pages={327},
  year={1966}
}

@article{PageP76,
  title={Multi-product inventory situations with one restriction},
  author={E. Page and Ray J. Paul},
  journal={Journal of the Operational Research Society},
  volume={27},
  number={4},
  pages={815--834},
  year={1976},
  publisher={Taylor \& Francis}
}

@article{GallegoSS92,
  title={The complexity of the staggering problem, and other classical inventory problems},
  author={Gallego, Guillermo and Shaw, Dong and Simchi-Levi, David},
  journal={Operations Research Letters},
  volume={12},
  number={1},
  pages={47--52},
  year={1992}
}

@article{Zoller77,
  title={Deterministic multi-item inventory systems with limited capacity},
  author={Zoller, Klaus},
  journal={Management Science},
  volume={24},
  number={4},
  pages={451--455},
  year={1977}
}

@book{Zipkin00,
  title={Foundations of Inventory Management},
  author={Zipkin, Paul Herbert},
  year={2000},
  publisher={McGraw-Hill}
}

@book{MuckstadtS10,
  title={Principles of Inventory Management: When You Are Down to Four, Order More},
  author={Muckstadt, John A. and Sapra, Amar},
  year={2010},
  publisher={Springer Science \& Business Media}
}

@article{GareyJ75,
  title={Complexity results for multiprocessor scheduling under resource constraints},
  author={Garey, Michael R. and Johnson, David S.},
  journal={SIAM Journal on Computing},
  volume={4},
  number={4},
  pages={397--411},
  year={1975}
}

@book{GareyJ79,
  title={Computers and Intractability},
  author={Garey, Michael R. and Johnson, David S},
  volume={174},
  year={1979},
  publisher={W. H. Freeman and Company}
}

\addtocontents{toc}{\protect\setcounter{tocdepth}{1}} 
\appendix
\section{Additional Proofs}

\subsection{Proof of Claim~\ref{clm:Cpstar_vs_M}} \label{app:proof_clm_Cpstar_vs_M}

Focusing on a single commodity $i \in [n]$, suppose we wish to identify a minimum-cost stationary order sizes and stationary intervals (SOSI) policy, subject to our warehouse constraint. Based on the discussion in Section~\ref{subsec:model_definition}, any such policy is representable by a single parameter, $T_i$, standing for the uniform time interval between successive orders. As such, its long-run average cost can be written as $C_{\myeoq, i}(T_i) = \frac{ K_i }{ T_i } + H_i T_i$, and due to attaining a peak inventory level of $T_i$, capacity-feasibility is equivalent to requiring that $\gamma_i T_i \leq {\cal V}$. Therefore, the setting we are considering can be succinctly formulated as:
\begin{equation} \label{eqn:proof_clm_Cpstar_vs_M_prob}
\begin{array}{lll}
{\displaystyle \min_{T_i}} & {\displaystyle C_{\myeoq, i}(T_i)} \\
\text{s.t.} & {\displaystyle  T_i \in (0,{\cal V}/\gamma_i]} 
\end{array}
\end{equation}
By consulting Claim~\ref{clm:EOQ_properties}, we know that $\sqrt{K_i/H_i}$ is the global minimizer of $C_{\myeoq, i}$ over $(0,\infty)$. However, since the above-mentioned search is limited to $(0,{\cal V}/\gamma_i]$, the convexity of this function implies that an optimal ordering interval for problem~\eqref{eqn:proof_clm_Cpstar_vs_M_prob} is given by $T_i^* = \min \{ \sqrt{K_i/H_i}, {\cal V}/ \gamma_i \}$; this term coincides with our definition of $T^{\cal V}_i$.

Now, to propose a single capacity-feasible replenishment policy for all commodities, we simply glue together the optimal capacity-feasible SOSI policies for the individual commodities, $T_1^*, \ldots, T_n^*$, and  scale down the combined policy by a factor of $n$ to make it capacity-feasible. According to Claim~\ref{clm:EOQ_properties}(3), this alteration increases the total cost by a factor of at most $n$, and since $C( {\cal P}^*)$ is the optimal long-run average cost, we have
\[ C( {\cal P}^* ) ~~\leq~~ n \cdot \sum_{i \in [n]} C_{\myeoq, i}( T_i^* ) ~~=~~ n \cdot \sum_{i \in [n]} \left( \frac{ K_i }{ T_i^* } + H_i T_i^* \right) ~=~~ n \cdot \sum_{i \in [n]} \left( \frac{ K_i }{ T^{\cal V}_i } + H_i T^{\cal V}_i \right) ~~=~~ n {\cal M} \ . \]

\subsection{Proof of Claim~\ref{clm:aux_holding_inequality}} \label{app:proof_clm_aux_holding_inequality}

We begin by observing that, since the policy $\tilde{\cal P}_i$ places $N( \tilde{\cal P}_i, [0,\tau_{\mycyc}))$ orders for commodity $i$, its holding cost ${\cal H}( \tilde{\cal P}_i, [0,\tau_{\mycyc}) )$ can be lower-bounded by that of the holding-cost-wise cheapest policy for this commodity across the cycle $[0,\tau_{\mycyc})$, subject to placing exactly $N( \tilde{\cal P}_i, [0,\tau_{\mycyc}))$ orders. In other words, our lower bound on ${\cal H}( \tilde{\cal P}_i, [0,\tau_{\mycyc}) )$ is given by the optimum value of:
\begin{equation} \label{eqn:proof_clm_aux_holding_inequality_prob}
\begin{array}{ll}
{\displaystyle \min_{{\cal P}_i}} & {\cal H}( {\cal P}_i, [0,\tau_{\mycyc}) ) \\
\text{s.t.} & N( {\cal P}_i, [0,\tau_{\mycyc})) = N( \tilde{\cal P}_i, [0,\tau_{\mycyc}))
\end{array}    
\end{equation}
In what follows, we will show that an optimal policy in this context places $N( \tilde{\cal P}_i, [0,\tau_{\mycyc}))$ evenly-spaced orders, each consisting of $\frac{ \tau_\mycyc }{ N( \tilde{\cal P}_i, [0,\tau_{\mycyc})) }$ units, leading to a total holding cost of $H_i \cdot \frac{ \tau_\mycyc^2 }{ N( \tilde{\cal P}_i, [0,\tau_{\mycyc})) }$. Consequently, for every frequency class $q \in [Q]$ and commodity $i \in \tilde{\cal F}_q$, we have
\[ {\cal H}( \tilde{\cal P}_i, [0,\tau_{\mycyc}) ) ~~\geq~~ H_i \cdot \frac{ \tau_\mycyc^2 }{ N( \tilde{\cal P}_i, [0,\tau_{\mycyc})) } ~~\geq~~ H_i \cdot \frac{ \tau_\mycyc^2 }{ (n/\eps)^{3q} } \ , \]
where the last inequality holds since $N( \tilde{\cal P}_i, [0,\tau_{\mycyc})) \leq (\frac{n}{\eps})^{3q}$, by definition of $\tilde{\cal F}_q$.

To establish the above-mentioned characterization of an optimal policy for problem~\eqref{eqn:proof_clm_aux_holding_inequality_prob}, let us focus on one such policy ${\cal P}_i$, and suppose that its $N_i = N( \tilde{\cal P}_i, [0,\tau_{\mycyc}))$ orders decompose the cycle $[0,\tau_{\mycyc})$ into subintervals of lengths $\Delta_1, \ldots, \Delta_{N_i}$. With this notation, the combined holding cost of ${\cal P}_i$ can be written as ${\cal H}( {\cal P}_i, [0,\tau_{\mycyc}) ) = H_i \cdot \sum_{\nu \in [{N}_i]} \Delta_{\nu}^2$. We claim that, subject to $\sum_{\nu \in [{N}_i]} \Delta_{\nu} = \tau_{\mycyc}$, the latter sum-of-squares is minimized by choosing $\Delta_1 = \cdots = \Delta_{{N}_i} = \frac{ \tau_{\mycyc} }{ {N}_i }$. Indeed, for any vector $\Delta = (\Delta_1, \ldots, \Delta_{{N}_i}) \in \bbR^{ {N}_i }_+$ with $\sum_{\nu \in [{N}_i]} \Delta_{\nu} = \tau_{\mycyc}$, by employing the Cauchy-Schwarz inequality, we have
\[ \tau_{\mycyc}^2 ~~=~~ \left( \sum_{\nu \in [{N}_i]} \Delta_{\nu} \right)^2 ~~=~~ \langle \Delta, \vec{1} \rangle^2 ~~\leq~~ \langle \Delta,\Delta \rangle \cdot \langle \vec{1}, \vec{1} \rangle ~~=~~ {N}_i \cdot \sum_{\nu \in [{N}_i]} \Delta_{\nu}^2 \ . \]
By rearranging, $\sum_{\nu \in [{N}_i]} {\Delta}_{\nu}^2 \geq \frac{ \tau_{\mycyc}^2 }{ {N}_i }$, and on the other hand, $\Delta_1 = \cdots = \Delta_{{N}_i} = \frac{ \tau_{\mycyc} }{ {N}_i }$ attains this bound.

\end{document}